\documentclass[onecolumn, journal, 11pt]{IEEEtran} 
\usepackage{setspace}
\usepackage{epsfig}
\usepackage{times}
\usepackage{float}
\usepackage{afterpage}
\usepackage{amsmath}
\usepackage{amstext}
\usepackage{amssymb,bm}
\usepackage{latexsym}
\usepackage{color}
\usepackage{graphicx}
\usepackage{amsmath}
\usepackage{amsthm}
\usepackage{graphicx}
\usepackage[center]{caption}
\usepackage{pstricks}
\usepackage{subcaption}
\usepackage{booktabs}
\usepackage{ulem}

\usepackage{geometry}
\allowdisplaybreaks

\newcommand{\gapVIone}{11.7}

\newtheorem{thm}{Theorem}

\theoremstyle{remark}
\newtheorem{rem}{Remark}
\theoremstyle{definition}

\begin{document}

\title{On the Capacity Region of the Two-user Interference Channel with a Cognitive Relay}

\author{Alex Dytso, Stefano Rini, Natasha Devroye and Daniela Tuninetti%
\thanks{%
Alex Dytso, Natasha Devroye, and Daniela Tuninetti are with the Electrical and Computer Engineering Department of the University of Illinois at Chicago, Chicago, IL 60607 USA (e-mail: \{odytso2, devroye, danielat\}@ uic.edu); their work was partially funded by NSF under award number 1017436; the contents of this article are solely the responsibility of the author and do not necessarily represent the official views of the NSF.
Stefano Rini was with Stanford University, Stanford, CA 94305 USA; he currently is  with the Department of Electrical and Computer Engineering National Chiao Tung University (NCTU) Hsinchu, Taiwan,  (e-mail: stefano@nctu.edu.tw); his work was partially funded by National Science Council under MOST 103-2218-E-009-014-MY2.
The results in this paper were presented in part in
the 2010 IEEE Information Theory Workshop \cite{rini:ITW20102} and
the 2012 IEEE International Conference on Communications \cite{LDIFC_CR}.
}}

\maketitle

\begin{abstract}
This paper considers a variation of the classical two-user interference channel where the communication of two interfering source-destination pairs  
is aided by an additional node 
that has a priori knowledge of the messages to be transmitted, which is referred to as the {\it cognitive relay}.
For this Interference Channel with a Cognitive Relay (ICCR) 
novel outer bounds and capacity region characterizations are derived.
In particular, for the class of injective semi-deterministic ICCRs, a sum-rate upper bound is derived for the general memoryless ICCR and further tightened for the Linear Deterministic Approximation (LDA) of the Gaussian noise channel at high SNR, which disregards the noise and focuses on the interaction among the users' signals.
The capacity region of the symmetric LDA is completely characterized 
{ except for the regime of moderately weak interference and weak links from the CR to the destinations.
The insights gained from the analysis of the LDA} are then translated back to the symmetric Gaussian noise channel (GICCR).
For the symmetric GICCR, an approximate characterization (to within a constant gap) of the capacity region is provided for a parameter regime where capacity was previously unknown.  The approximately optimal scheme suggests that message cognition at a relay is beneficial for interference management as it enables simultaneous {\it over the air neutralization} of the interference at both destinations.
\end{abstract}


\section{Introduction}
\begin{figure}
\centering
        \begin{subfigure}[b]{0.30\textwidth}
                \includegraphics[width=\textwidth]{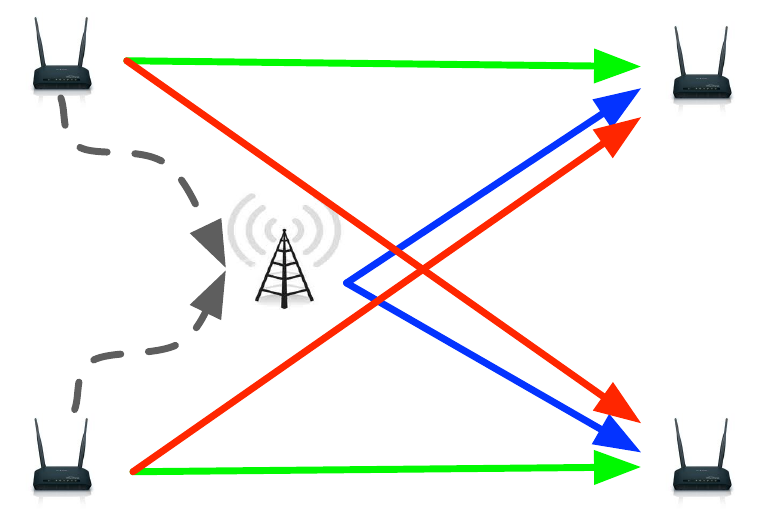}
                \caption{\small Network aided by a small cell.}
                \label{fig:3models_a}
        \end{subfigure}
\hspace*{1cm}
        \begin{subfigure}[b]{0.40\textwidth}
                \includegraphics[width=\textwidth]{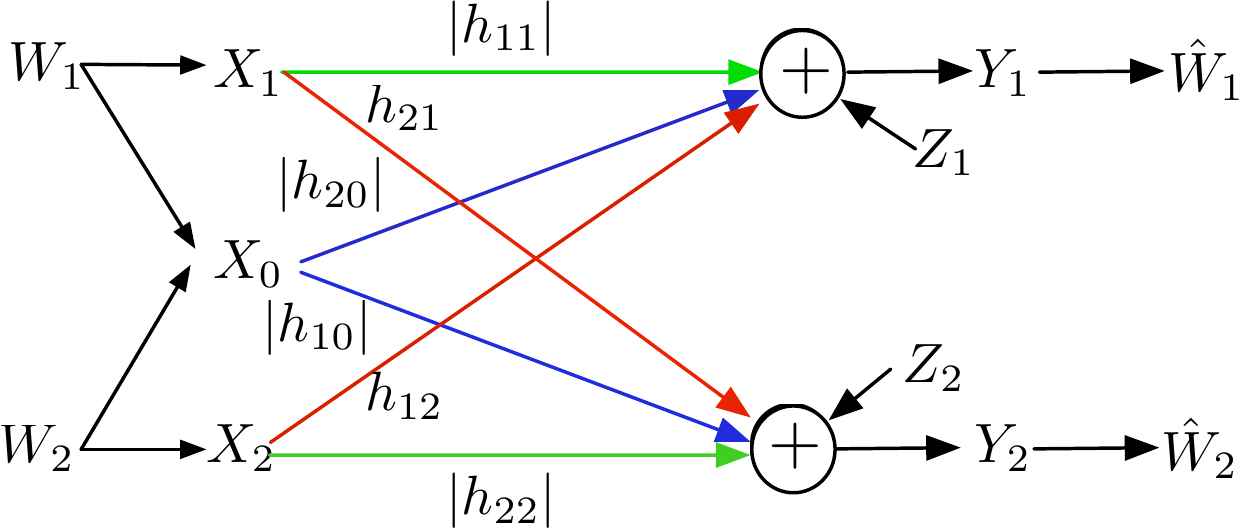}
                \caption{\small The two-user GICCR.}
                \label{fig:3models_b}
        \end{subfigure}
\caption{\small Network model under investigation.}
\label{fig:3models}
\end{figure}

In the last two decades
the wireless industry has grown at such a rapid rate that it has started to exhaust much of the precious frequency spectrum \cite{FCC-crunch}.
As a response, new technologies have emerged to improve spectrum management. 
Pico and femto cells technologies \cite{pico-TP, pico-primer}, for example, 
use many small base-stations with relatively small coverage areas (as opposed to and in addition to standard macro base-stations with larger coverage areas) 
to achieve higher throughputs through aggressive spatial reuse.
When the small cells have knowledge of the messages to be transmitted by the macro base-stations, they may act as relays to help other devices on the network, as shown in Fig.~\ref{fig:3models_a}, 
by providing an additional communication path for a message to the desired destination, and 
by allowing the small cell to better manage / combat the interference.
In this work we seek to obtain insights into the performance of such small cell inspired systems.
We take an information theoretic approach to the study of such architectures in order to obtain technology-independent characterizations on the possible performance  of the system, measured here in terms of capacity regions.

We study 
the Interference Channel with a Cognitive Relay (ICCR) shown in Fig.~\ref{fig:3models_b}.
In this simple model, the ICCR has two independent sources (macro base stations) that send information to their respective destinations by sharing the same channel, i.e., interfering with each other. In addition, a relay (small cell base station) that is non-causally aware of both messages before transmission starts, aids the two sources.  Since the relay knows both messages,  we term it the {\it  Cognitive Relay} (CR) following \cite{devroye_IEEE}.  Non-causal message knowledge at the relay may be possible when the relay
backhauls to the other transmitting nodes. 
Alternatively, if no backhauls are possible, the relay may learn the messages of the other transmitters over the air in a causal fashion---in this case the  model studied may provide a useful outer bound to the performance of any causal system.
Non-causal message knowledge could also be the result of a failed transmission in systems employing retransmission protocols.
Besides its practical motivations, the ICCR 
is also independently interesting from a 
theoretical perspective as it generalizes several channel models: the Interference Channel (IC) \cite{Meulen:1977}, when the relay is not present, the Broadcast Channel (BC) \cite{cover_BC_1977}, when both transmitters are not present, and the Cognitive Interference Channel (CIC), when one source is not present \cite{devroye_IEEE}.

\paragraph*{Past Work}
In this work we focus on the case where the relay has {\it  non-causal message knowledge} and is {\it  in-band}, that is, the CR shares the same channel as the two source-destination pairs.  We note however that significant work exists on models with causal cognition at the relay (where the relay is in-band \cite{Sahin_2007_1}, or out-of-band \cite{sahin:ITA:2009}, or out-of-band with noiseless rate-limited links from the CR to the destinations \cite{razaghi2011two}, and others variations such as those investigated in \cite{BoundPotentRelayTianYener,GIFoutOfBandRelayTianYener}).
If only one message is available non-causally we obtain a MIMO CIC studied in \cite{Rini:ISIT2013}. Finally, if only portions of the messages are available, the techniques developed in this paper would apply to the portion of the messages known at the relay, but decoding rates would differ (as for example, partial message knowledge does not allow for complete neutralization of the interference) and might more resemble that of an IC.

To the best of our knowledge, the 
ICCR was first considered in \cite{Sahin_2007_2}, where an achievable rate region was proposed. This rate region was improved upon in Gaussian noise in \cite{sridharan2008capacity}, and again for a general discrete memoryless channel in \cite{jiang-achievable-BCCR}.
%
The authors of \cite{sridharan2008capacity} first proposed a sum-rate outer bound for the Gaussian channel.
In our conference work \cite{rini:ITW20102}, we derived the first outer bound for a general memoryless channel, which we further tightened for a class of semi-deterministic channels subject to injectivity conditions in the spirit of \cite{telatar2008bounds}. The tightened outer bound was shown in  \cite{rini:ITW20102} to be capacity for the class of Linear Deterministic Approximation (LDA) of the Gaussian noise channel at high SNR (first introduced in \cite{avestimehr2011wireless})
in the absence of interfering links and in several other special cases.
In \cite{rini:isit2011:strong,riniIFCCR} general inner and outer bounds were obtained, and shown to match for a class of ICCRs with ``very strong interference at one destination.''
In our conference work \cite{LDIFC_CR}, the capacity of the symmetric ICCR LDA was shown for almost all channel parameters with a tighter outer bound than that in \cite{rini:ITW20102}.  The insight from the capacity achieving schemes were used to show capacity to within 3 bits/sec/Hz in the Gaussian ICCR (GICCR) without interfering links in \cite{rini:isit2011:3}, which was recently improved upon by \cite{PC-CRcapacity}, where capacity was shown for this channel through the derivation of a new outer bound tailored to the channel without interfering links.

\paragraph*{Contributions}
The general ICCR is a complex channel model that generalizes 
many classical channel models for which capacity is open, 
including the IC and the BC. As such, deriving its capacity region  is a challenging and ambitious task. We approach this task by first focusing our attention on the LDA, which highlights the interplay between users' signals by eliminating the randomness of the noise \cite{avestimehr2011wireless}. The LDA models the Gaussian channel at high SNR, and as such, schemes developed for the LDA can often be translated into ``good'' achievable strategies for the Gaussian noise channel at any finite SNR, 
that is, although not optimal in the sense of exactly achieving an outer bound, they lie within a bounded distance of the outer bounds regardless of the channel parameters. 
This approach has allowed for progress on long standing open problems; for example, the capacity of the IC \cite{etkin_tse_wang} and of the CIC \cite{rini:journal2} are known to within 1~bit/sec/Hz. In this work, we first analyze the symmetric LDA
by determining its capacity region in almost all parameter regimes (roughly speaking, the case of weak links from the CR to the destinations and of moderately weak interference at a destination from the non-intended source is excluded). We present new achievable techniques that are sum-capacity optimal for the LDA model that were not presented in our conference work  \cite{LDIFC_CR}.
Then, with the insight gained from the study of the LDA, we move back to the symmetric Gaussian noise channel and show capacity to within a constant gap in several parameter regimes that mimic the capacity results for the LDA, which has not appeared in any conference version of our work.

Our central contributions are: 
(1) Deriving novel (non cut-set) outer bounds for the class of injective semi-deterministic ICCRs;
(2) Further tailoring and tightening of the outer bounds for the LDA and GICCR;
(3) Deriving optimal achievability schemes in almost all parameter regimes for the symmetric LDA  
and providing insight into what might be missing in the parameter regime for which we do not have capacity; 
(4) Deriving the capacity to within a constant gap for the symmetric Gaussian channel in regimes where it was open; 
(5) Numerically comparing the proposed inner and outer bounds with other achievability schemes.

We note that the central contribution of this work lies in considering a {\it general} ICCR for the outer bound, rather than models where the assumptions of strong interference \cite{riniIFCCR} or the absence of interfering links \cite{PC-CRcapacity} significantly simplify the problem. For sake of space, and in order to convey the key contributions of this work, we focus here only on symmetric channels for achievability results, that is, ICCRs in which the capacity is the same when the role of the sources is swapped. This is done so as to reduce the number of parameters
and obtain insightful analytically tractable results. Nevertheless, our outer bounds and achievable scheme apply to general non-symmetric LDAs and GICCRs. We expect that the extension of the presented analysis to the fully general ICCR may follow the same approach used here for the symmetric case, albeit with more involved analytical computations.

\paragraph*{Paper Organization}
We introduce the channel model in Section \ref{sec:CH}.
In Section \ref{sec:Bounds} we present our novel outer bounds. 
In Section \ref{sec:Cap:LDA} we determine the capacity region for the LDA in almost all parameter regimes.
In Section \ref{sec:Gaussian} we derive the capacity to within a constant gap for some parameter regimes of the Gaussian channel which were open, and numerically compare the inner and outer bounds with other relaying schemes.
Section \ref{sec:conclusion} concludes the paper.
Some proofs are found in the Appendix.

\paragraph*{Notation}
We use the notation convention of \cite{elgamalkimbook}:
$[n_1:n_2]$ is the set of integers from $n_1$ to $n_2\geq n_1$;
$[x]^+ := \max\{0,x\}$ for $x\in\mathbb{R}$;
$x^n$ denotes a vector of length $n$ with components $(x_1,\ldots,x_n)$;
lower case $x$ is an outcome of random variable in upper case $X$ which lies in calligraphic case alphabet $\mathcal{X}$;
$\mathcal{N}(\mu,\sigma^2)$ denotes a proper-complex Gaussian random variable with mean $\mu \in\mathbb{C}$ and variance $\sigma^2\in\mathbb{R}_+$;
$\delta(\cdot)$ denotes the Dirac delta function.

\section{Channel Models}
\label{sec:CH}

\subsection{The General Memoryless ICCR}

The general two-user memoryless ICCR 
is characterized by three input alphabets $(\mathcal{X}_0,\mathcal{X}_1,\mathcal{X}_2)$,
two output alphabets $(\mathcal{Y}_1,\mathcal{Y}_2)$, and a memoryless transition probability $\mathbb{P}_{Y_1,Y_2|X_0,X_1,X_2}$.
%
Source $i$, $i \in [1:2]$, encodes the message $W_i$, assumed independent of everything else and uniformly distributed on $[1:2^{n R_i}]$, into a codeword $X_i^n\in \mathcal{X}_i^n$, where $n\in\mathbb{N}$ denotes the codeword length and $R_i$ the rate in bits per channel use.
Message $W_i$ is intended for receiver $i$, $i \in [1:2]$, which forms the estimate $\widehat{W}_i$ from channel output $Y_i^n \in \mathcal{Y}^n_i$.
The two sources are aided by a cognitive relay that has knowledge of and encodes $W_1$ and $W_2$ into the codeword $X_0^n \in \mathcal{X}_0^n$.
A non-negative rate pair $(R_1,R_2)$ is said to be achievable if there exists a sequence of encoding functions
$
X_1^n(W_1), \
X_2^n(W_2), \
X_0^n(W_1,W_2),
$
and decoding functions
$
\widehat{W}_1(Y_1^n), \
\widehat{W}_2(Y_2^n),
$

such that  the maximum probability of error satisfies
$\max_{i \in [1:2]}\mathbb{P}[\widehat{W}_i \neq W_i] \rightarrow 0$ as $n \rightarrow +\infty.$
The capacity region is the convex closure of all achievable rate pairs $(R_1,R_2)$ \cite{elgamalkimbook}.

Since the destinations do not cooperate, the channel capacity only depends on the conditional marginal distributions $\mathbb{P}_{Y_\ell|X_0,X_1,X_2}(y_\ell|x_0, x_1,x_2)$, $\ell\in[1:2]$.
In other words, all ICCRs that share the same conditional marginal distributions have the same capacity region, as for the BC \cite[Lemma 5.1]{elgamalkimbook}. 
Note that the ICCR contains three important channels as special cases:
(a) the IC if $X_0= \emptyset$,
(b) the BC if $X_1= X_2= \emptyset$ ,
(c) the CIC if either $X_1= \emptyset$ or $X_2= \emptyset$.

\subsection{The Injective Semi-deterministic ICCR}
\label{sec:CH:inject}

The injective semi-deterministic ICCR was introduced in \cite{Telatar_Tse_inject} and corresponds to the special case when the transition probability satisfies
\begin{align}
&\mathbb{P}_{Y_1,Y_2|X_0,X_1,X_2}(y_1,y_2|x_0,x_1,x_2) \notag\\
&= \sum_{v_1,v_2}\mathbb{P}_{V_1|X_1}(v_1|x_1)\mathbb{P}_{V_2|X_2}(v_2|x_2)
\ \delta\Big(y_1 - f_{1}(x_1,x_0,v_2)\Big)
\ \delta\Big(y_2 - f_{2}(x_2,x_0,v_1)\Big),
\label{eq:def of semidet inj gen}
\end{align}
for some memoryless transition probabilities $\mathbb{P}_{V_1|X_1}$ and $\mathbb{P}_{V_2|X_2}$, and some deterministic functions $f_{1}$ and $f_{2}$ that are injective when $(X_1,X_0)$ and $(X_2,X_0)$,  respectively, are held fixed, which implies that for all $\mathbb{P}_{X_0, X_1,X_2}$ one has
$H(Y_1|X_1,X_0)=H(V_2|X_1,X_0)=H(V_2|X_0)$,
and similarly for the other source.
Injective semi-deterministic channels are important because 
approximate capacity results for this class of channels are available while those of their more general counterpart are still open. For example, the injective deterministic IC (where $\mathbb{P}_{V_1|X_1}$ and $\mathbb{P}_{V_2|X_2}$ are noiseless) was completely solved in \cite{elgamal_det_IC} and the injective semi-deterministic IC capacity  was characterized to within a constant gap in \cite{telatar2008bounds}. Intuitively, it is easier to characterize the capacity of an injective semi-deterministic IC, compared to the general IC, as one knows exactly what the interference signals are through the random variables $V_1$ and $V_2$ \cite{telatar2008bounds}. We also note that the important Gaussian channel is a special case of the  injective semi-deterministic model. For continuous alphabets, the summations in~\eqref{eq:def of semidet inj gen} must be replaced with integrals.

\subsection{The GICCR}
The complex-valued single-antenna power-constrained GICCR in a standard form \cite{riniIFCCR} is shown in Fig.~\ref{fig:3models_b} and is described by the input-output relationships
\begin{subequations}
\begin{align}
Y_1 &= |h_{11}|X_1+|h_{10}|X_0+ h_{12} X_2+Z_1, \\
Y_2 &=  h_{21} X_1+|h_{20}|X_0+|h_{22}|X_2+Z_2,
\end{align}
\label{eq:GICCR}
\end{subequations}
where, without loss of generality,  the input $X_i \in \mathbb{C}$ is subject to power constraint $\mathbb{E}[|X_i|^2] \leq 1$, $i\in[0:2]$,
the noise $Z_j \sim \mathcal{N}(0,1)$, $j\in[1:2]$, and
%
the channel gains $h_{ij}$, $i\in[1:2], j\in[0:2]$, are complex-valued, fixed and known to all nodes.
Without loss of generality, some channel gains can be taken to be real-valued and non-negative
\cite[Appendix M]{riniIFCCR}.
The Gaussian GICCR is a special case of the injective semi-deterministic ICCR in~\eqref{eq:def of semidet inj gen}, where
$V_1 := h_{21} X_1 + Z_2$,
$V_2 := h_{12} X_2 + Z_1$,
and $f_{1}$ and $f_{2}$ are complex-valued linear combinations.

The capacity region of the GICCR is open. Progress towards understanding its fundamental limits is possible by providing an approximate characterization of its capacity as pioneered in \cite{avestimehr2011wireless}.
The capacity is said to be known {\it  to within $\mathsf{GAP}$~bits} if one can show an inner bound region $\mathcal{I}$ and an outer bound region $\mathcal{O}$ such that { $(R_1,R_2)\in \mathcal{O} \Longrightarrow ([R_1-\mathsf{GAP}]^+,[R_2-\mathsf{GAP}]^+)\in \mathcal{I}$}.
The $\mathsf{GAP}$ upper bounds the worst-case distance between the inner and outer bounds \cite{etkin_tse_wang}. 
%

\subsection{The LDA}
The {\it  linear deterministic approximation} (LDA) of the GICCR is a model that captures the behavior of the GICCR in~\eqref{eq:GICCR} at high SNR. The LDA is a fully deterministic model described by \cite{avestimehr2011wireless} 
\begin{subequations}
\begin{align}
Y_1 &=  \mathbf{S}^{m-n_{11}} X_1
 \oplus \mathbf{S}^{m-n_{10}} X_0
 \oplus \mathbf{S}^{m-n_{12}} X_2, \\
Y_2 &=  \mathbf{S}^{m-n_{21}} X_1
 \oplus \mathbf{S}^{m-n_{20}} X_0
 \oplus \mathbf{S}^{m-n_{22}} X_2,
\end{align}
\label{eq:LDICCR}
\end{subequations}
where $\mathbf{S}$ is the binary down-shift matrix of dimension
$m:=\max\{n_{11},n_{12},n_{10},n_{21},n_{22},n_{20}\}$, for $\{n_{ij}\in \mathbb{N},  \ i\in[1:2], j\in[0:2]\}$,
all inputs and outputs are binary column vectors of dimension $m$,
and where the symbol $\oplus$ denotes the component-wise modulo-2 addition of the binary vectors.
The LDA is a special case of injective semi-deterministic ICCR in~\eqref{eq:def of semidet inj gen} where
$V_1 := \mathbf{S}^{m-n_{21}} X_1$,
$V_2 := \mathbf{S}^{m-n_{12}} X_2$,
and $f_{1}$ and $f_{2}$ are modulo-2 additions.
The LDA in~\eqref{eq:LDICCR} may be related to the GICCR in~\eqref{eq:GICCR} by taking $n_{ij} = \lfloor \log(1+|h_{ij}|^2)\rfloor$ \cite{etkin_tse_wang}.
The capacity of the LDA often gives insight into strategies that are optimal to within a constant  gap for the GICCR~\cite{bresler_tse}.

\section{Outer Bounds}
\label{sec:Bounds}

We start off stating a known outer bound  for the general memoryless ICCR, and then deriving new outer bounds for the injective semi-deterministic ICCR, the LDA and the GICCR.
\begin{thm}{\it (Outer bound to the capacity of the general memoryless ICCR \cite[Thm. III.1]{riniIFCCR}).}
\label{thm:outer bound:DM}
If $(R_1,R_2)$ lies in the capacity region of the general memoryless ICCR, then
\begin{subequations}
\begin{align}
R_1     &\leq  I(Y_1;X_1,X_0|Q,X_2), \label{eq:out:DM1}\\
R_2     &\leq  I(Y_2;X_2,X_0|Q,X_1), \label{eq:out:DM2}\\
R_1+R_2 &\leq  I(Y_2;X_1,X_2,X_0|Q)
+ I(Y_1;X_1,X_0|Q,\bar{Y}_2,X_2), \label{eq:out:DM3}\\
R_1+R_2 &\leq  I(Y_1;X_1,X_2,X_0|Q)
+ I(Y_2;X_2,X_0|Q,\bar{Y}_1,X_1), \label{eq:out:DM4}
\end{align}
\label{eq:outer bound:DM}
\end{subequations}
for some input distribution that factors as
$\mathbb{P}_{Q,X_1,X_2,X_0}=
\mathbb{P}_Q
\mathbb{P}_{X_1|Q}
\mathbb{P}_{X_2|Q}
\mathbb{P}_{X_0|X_1,X_2,Q}
$,
where $\bar{Y}_1$ and $\bar{Y}_2$ 
have the same conditional marginal distributions of the channel outputs $Y_1$ and $Y_2$ given the inputs $(X_1,X_2,X_0)$, respectively, but are otherwise arbitrarily jointly distributed.
\hfill$\square$\end{thm}

The region in Theorem~\ref{thm:outer bound:DM} is not the tightest in general. For example, \cite{riniIFCCR} reports other outer bounds that can actually be used to prove capacity in some regimes. However, these other outer bounds depend on auxiliary random variables for which no cardinality bound is known on the corresponding alphabets. The advantage of Theorem~\ref{thm:outer bound:DM} is that it only contains random variables defined in the channel model (with the exception of the time-sharing random variable $Q$) and it is therefore in principle computable.
Note that the correlation among $\bar{Y}_1$ and $\bar{Y}_2$ in Theorem~\ref{thm:outer bound:DM} may be chosen to tighten the bound since the capacity region of the ICCR is only a function of the output conditional marginal distributions, as for the BC \cite[Lemma 5.1]{elgamalkimbook} and the CIC~\cite{riniIFCCR}.


Theorem~\ref{thm:outer bound:DM} reduces to:
(a) the capacity region of a deterministic BC when $X_1=X_2=\emptyset$ \cite{elgamalkimbook}, and
(b) the capacity region of a deterministic CIC when either $X_2=\emptyset$ or $X_1=\emptyset$ \cite{rini:journal1}.
However, it does not reduce to the capacity region of the class of fully deterministic IC when $X_0=\emptyset$ \cite{elgamal_det_IC}.
Hence, in the following we develop additional rate bounds that reduce to the bounds for the injective semi-deterministic IC developed in \cite{telatar_tse}, which includes the fully deterministic IC studied in \cite{elgamal_det_IC}, when $X_0=\emptyset$.

\subsection{Novel Outer bounds for the Injective Semi-deterministic ICCR}
The outer bound of Theorem~\ref{thm:outer bound:DM} may be tightened for the injective semi-deterministic ICCR defined in~\eqref{eq:def of semidet inj gen} as follows, whose proof can be found in the Appendix~\ref{proof{thm:outer bound:inj}}:
\begin{thm}
\label{thm:outer bound:inj}
If $(R_1,R_2)$ lies in the capacity region of the injective semi-deterministic ICCR, then in addition to the bounds in~\eqref{eq:outer bound:DM}, the following must hold
\begin{subequations}
\begin{align}
R_1+R_2
  &\leq  H(Y_1|\widetilde{V}_1,Q)-H(\widetilde{V}_2|X_2)
    +    H(Y_2|\widetilde{V}_2,Q)-H(\widetilde{V}_1|X_1)
    +\mathsf{MLP}_1,
    \label{eq:outer bound:inj sumrate}
\\
2R_1+R_2
  &\leq -H(\widetilde{V}_1|X_1)-2H(V_2|X_2)
     + H(Y_1|Q)
     + H(Y_1|\widetilde{V}_1,X_2,Q)
     + H(Y_2|\widetilde{V}_2,Q)
     +\mathsf{MLP}_1,
    \label{eq:outer bound:inj sumrate 2r1+r2}
\\
R_1+2R_2
  &\leq -H(\widetilde{V}_2|X_1)-2H(V_1|X_2)
    + H(Y_2|Q)
    + H(Y_2|\widetilde{V}_2,X_1,Q)
    + H(Y_1|\widetilde{V}_1,Q)
    +\mathsf{MLP}_1,
    \label{eq:outer bound:inj sumrate r1+2r2}
\end{align}
where the multi-letter portion (MLP) $\mathsf{MLP}_1$ is given by
\begin{align}
\mathsf{MLP}_1 &:= {\sup_{n\in\mathbb{N}}} \ \frac{1}{n}\Big(
 I(V_1^n;X_0^n { | W_2})
+I(V_2^n;X_0^n { | W_1})
\Big),
\label{eq:outer bound:inj:mlp1}
\end{align}
and where the random variables $\widetilde{V}_1$ and $\widetilde{V}_2$ are conditionally independent copies of $V_1$ and $V_2$, respectively, that is, they are jointly distributed with $(Q,X_1,X_2,X_0)$ as
\begin{align}
\mathbb{P}_{\widetilde{V}_{1},\widetilde{V}_{2}|Q,X_1,X_2,X_0}(v_1,v_2|q,x_1,x_2,x_0)
=
\mathbb{P}_{V_1|X_1}(v_1|x_1)
\mathbb{P}_{V_2|X_2}(v_2|x_2),
\end{align}
where $\mathbb{P}_{V_1|X_1}$ and $\mathbb{P}_{V_2|X_2}$ are part of the channel model definition in~\eqref{eq:def of semidet inj gen}.
\label{eq:outer bound:inj}
\end{subequations}
\hfill$\square$\end{thm}

The auxiliary random variables $\widetilde{V}_1$ and $\widetilde{V}_2$ are provided as ``genie side information'' at receivers 1 and 2, respectively, as a mathematical tool to enable the derivation of ``single letter'' outer bounds;
they are identical to those used in \cite{telatar_tse}, and thus with this choice
Theorem~\ref{thm:outer bound:inj} reduces to \cite[Theorem~1]{telatar_tse}, which is tight  for the LDA \cite{bresler_tse} and is optimal to within 1~bit for the Gaussian IC \cite{etkin_tse_wang}, when $X_0=\emptyset$.
%
Theorem~\ref{thm:outer bound:inj} is however not in the  desirable ``single-letter'' format,  as  it contains the MLP in~\eqref{eq:outer bound:inj:mlp1}. We discuss how to ``single-letterized'' the MLP in~\eqref{eq:outer bound:inj:mlp1} for the LDA and the GICCR in the rest of the section.

{
Note that the step of tightening the bound used in the proof of
Theorem~\ref{thm:outer bound:inj}
(i.e., conditioning on the interference function $V_j$ rather then on the interfering codeword $X_j, \ j\in[1,2]$)
highlights a stumbling block in deriving outer bounds for general IC and BC: in general
we do not know the exact form of the interfering signal(s) at
a receiver for any possible input distribution. Assuming that
the channel is deterministic and in a certain way invertible,
allows one to exactly determine the interference.
Notice that ``conditioning'' on the interference functions
$V_1$ or $V_2$ 
may be interpreted as if the interference has been removed without necessarily decoding the
corresponding messages. 
On the other hand, conditioning on $X_1$ or $X_2$
may be interpreted as if the message carried by $X_1$ or $X_2$ were known through decoding.
}

\subsection{Outer Bounds for the LDA}
For a discrete-valued channel (for which the entropy is non-negative), one may turn the MLP in~\eqref{eq:outer bound:inj:mlp1} into a single-letter expression as
\begin{align}
\mathsf{MLP}_1
&\leq {\sup_{n\in\mathbb{N}}} \  \frac{1}{n}\Big(
 \min\{H(V_2^n),H(X_0^n)\}
+\min\{H(V_1^n),H(X_0^n)\}
\Big)
\notag\\&
\leq
 \min\{H(V_2),H(X_0)\}
+\min\{H(V_1),H(X_0)\}.
\label{eq:mlp:singleletter:attempt2}
\end{align}
For the LDA we next provide a tighter bound than that in~\eqref{eq:mlp:singleletter:attempt2}.
The LDA belongs to a special class of injective deterministic channels whose outputs are described by
\begin{subequations}
\begin{align}
Y_1 = f_1\big(X_1,q_1(X_0),V_2\big), \ V_{2}=g_2\big(X_2\big), \quad
Y_2 = f_2\big(X_1,q_2(X_0),V_1\big), \ V_{1}=g_1\big(X_1\big),
\end{align}
\label{eq:further restriction of injective semidet to fit LDA}
\end{subequations}
where $q_1,q_2,g_1,g_2,f_1,f_2$ are deterministic functions.
The difference between~\eqref{eq:further restriction of injective semidet to fit LDA} and~\eqref{eq:def of semidet inj gen} is that in the former the output at receiver $i\in[1:2]$ depends on a function $q_i(X_0)$ rather than on $X_0$; this distinction is important when the function $q_i(\cdot)$ is not a bijection, as it may be the case in the LDA.
We further require the function $f_1$ to be injective when its first two arguments are known, that is, $H(Y_1|q_1(X_0),X_1) = H(V_2|q_1(X_0))$,
and analogously for $f_2$.

For the LDA, Theorem~\ref{thm:outer bound:inj} may be tightened as follows, whose proof may be found in Appendix~\ref{proof{thm:outer bound:LDA}}:
\begin{thm}
\label{thm:outer bound:LDA}
For the LDA the term $\mathsf{MLP}_1$ in~\eqref{eq:outer bound:inj:mlp1} can be tighten by using instead
\begin{align}
\mathsf{MLP}_2 &:= \min\{n_{20},n_{21}\}+\min\{n_{10},n_{12}\},
\label{eq:outer bound:LD:mlp}
\end{align}
and the resulting region in~\eqref{eq:outer bound:inj}
is exhausted by considering i.i.d.
Bernoulli($1/2$) inputs. 
\hfill$\square$\end{thm}


{
\subsection{Outer Bounds for the GICCR}

The proof of Theorem~\ref{thm:outer bound:LDA} inspired the following bound on the MLP for the Gaussian noise channel.
For the GICCR, Theorem~\ref{thm:outer bound:inj} may be tightened as follows, whose proof
may be found in Appendix~\ref{proof{thm:outer bound:Gaussian}}:
\begin{thm}
\label{thm:outer bound:Gaussian}
For the GICCR the term $\mathsf{MLP}_1$ in~\eqref{eq:outer bound:inj:mlp1} can be tighten by using instead
\begin{align}
\mathsf{MLP}_2 &:=
 \log(1+\min\{|h_{12}|^2,|h_{10}|^2\})
+\log(1+\min\{|h_{21}|^2,|h_{20}|^2\})
+2\log(2),
\label{eq:outer bound:Gaussian:mlp}
\end{align}
and the resulting region in~\eqref{eq:outer bound:inj}
is exhausted by considering jointly Gaussian inputs.
\hfill$\square$\end{thm}
}

\section{Capacity for the symmetric LDA}
\label{sec:Cap:LDA}

\begin{figure}
\centering
\includegraphics[width=12cm]{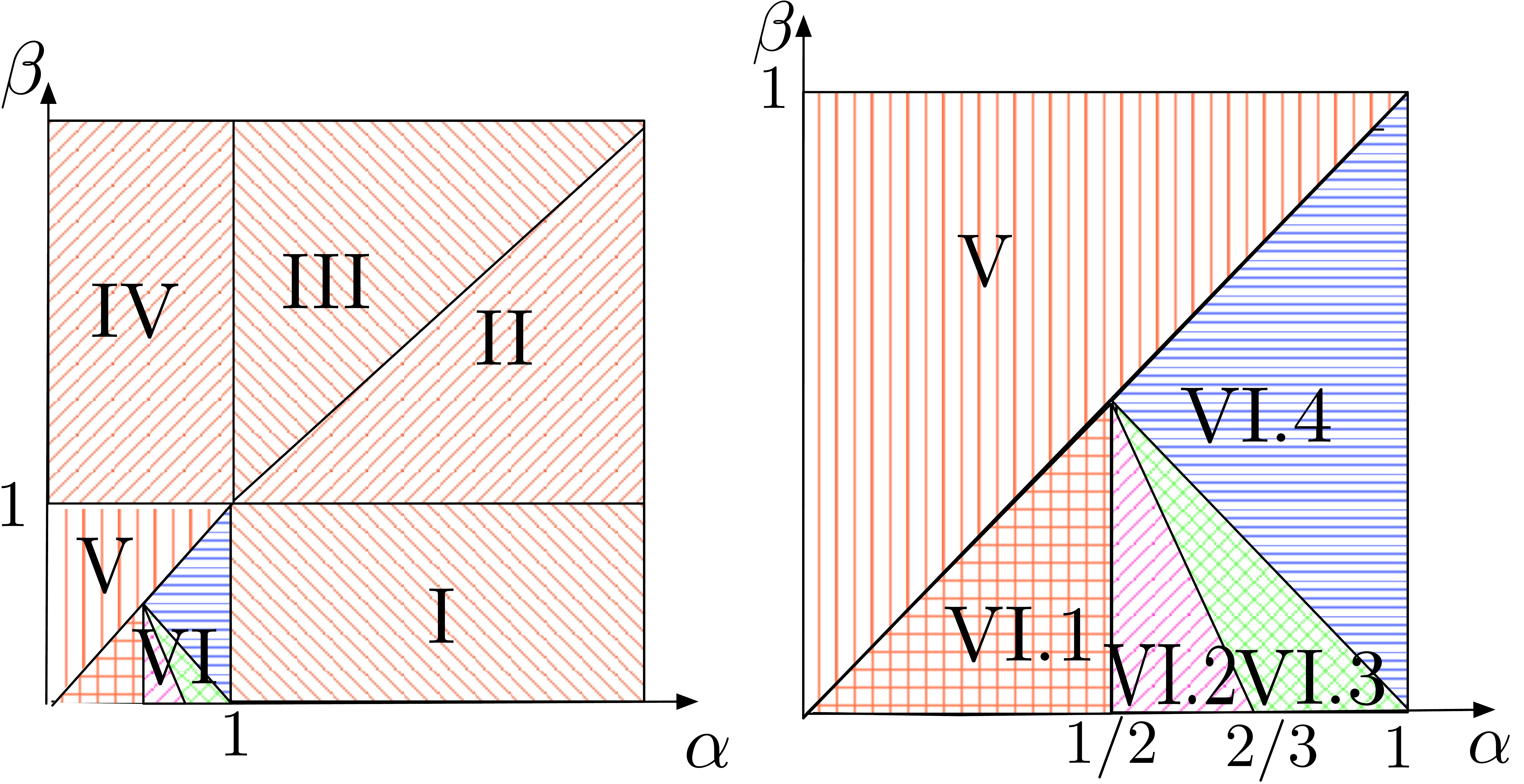}
\caption{\small
Parameter regimes for the LDA and the GICCR at high SNR.
Regimes I to VI.1: capacity is known for the LDA and  to within a constant gap for the GICCR.
Regimes VI.2 to VI.4: capacity is open for both models (but sum-capacity is known in some cases for the LDA).
}
\label{fig:Regions}
\end{figure}

In this section we propose achievable schemes that match the outer bound in Theorem~\ref{thm:outer bound:LDA} for almost all channel parameters, where channel gains and the rates are parametrized as
\begin{subequations}
\begin{align}
n_{11} &= n_{22} =n_{\rm{S}} > 0,\\
n_{12} &= n_{21} =n_{\rm{I}} =\alpha \ n_{\rm{S}}, \quad \alpha \geq 0,\\
n_{10} &= n_{20} =n_{\rm{C}} =\beta  \ n_{\rm{S}}, \quad \beta  \geq 0,\\
R_i   &=r_i \ n_{\rm{S}}, \quad r_i\geq 0, \quad i\in[1:2].
\end{align}
\label{eq:parameter LDA}
\end{subequations}
{The focus on the symmetric case is not for lack of generality of the developed theory but for simplicity of exposition (the symmetric model is specified by three parameters rather than six).}

Under the symmetric condition in~\eqref{eq:parameter LDA}, 
the outer bound in Theorem~\ref{thm:outer bound:LDA} 
simplifies to
\begin{subequations}
\begin{align}
r_1 &\leq  \max\{1, \beta\}, \quad 
r_2  \leq  \max\{1, \beta\}, \label{eq:DT OK SU2}\\
r_1+r_2 &\leq [1-\max\{\alpha,\beta\}]^+ +\beta + \max\{1,\alpha\}, \label{eq:DT OK KRA}\\ 
r_1+r_2 &\leq  \max\{1, \beta\}, \ \ \text{apply for $\alpha=1$ only}, \label{eq:DT OK TDMA}\\ 
r_1+r_2 &\leq 2\max\{1-\alpha,\alpha,\beta\} + 2\min\{\alpha,\beta\}, \label{eq:DT OK ETW} \\
2r_1+ r_2 &\leq \max\{1,\beta,\alpha\}+\max\{1-\alpha,\alpha,\beta\}
+ \max\{1-\alpha,\beta\}+2\min\{\alpha,\beta\}, \label{eq:DT OK 12}\\
 r_1+2r_2 &\leq \max\{1,\beta,\alpha\}+\max\{1-\alpha,\alpha,\beta\}
 + \max\{1-\alpha,\beta\}+2\min\{\alpha,\beta\}. \label{eq:DT OK 21}
\end{align}
\label{eq:highsnrbounds:sym}
\end{subequations}
%
The outer bound in~\eqref{eq:highsnrbounds:sym} naturally leads to the division of the channel parameter space $(\alpha,\beta)$ in \eqref{eq:parameter LDA} into the six regimes illustrated in Fig.~\ref{fig:Regions} based on the different values of the $\max$ / $\min$ terms in~\eqref{eq:highsnrbounds:sym}.

\begin{thm}
\label{thm:LDA:cap}
For the symmetric LDA, capacity is known for the following regimes (see Fig.~\ref{fig:Regions}):
Regimes I to V ($1\leq \max\{\alpha,\beta\}$) and
Regime VI.1 ($\beta \le \alpha \le \frac{1}{2}$).
For the remaining regimes the sum-capacity is known for $4\alpha -3 \leq \beta  \le 3\alpha -2, \  2/3 \leq \alpha \leq 1$, which in Regime VI.4 implies that the whole capacity region is known.
\hfill$\square$\end{thm}
\begin{IEEEproof}
We now prove Theorem~\ref{thm:LDA:cap} for different cases and regimes.

\setcounter{paragraph}{0}
\paragraph{Case $n_{\rm{S}}>0$ and $\alpha=1$ (line $\alpha=1$ in Fig. \ref{fig:Regions})}
The outer bound in~\eqref{eq:highsnrbounds:sym} when $\alpha=1$ is simply the triangle  formed by
$r_1 + r_2 \leq \max\{1,\beta\}$,
from~\eqref{eq:DT OK TDMA} only, which is trivially achieved by time division between the cases where one source is silent and the cognitive relay fully helps the other source. 
In particular, in order to prove capacity, it suffices to show the achievability of
$(r_1,r_2)=(\max\{1,\beta\},0)$, which can be attained as follows.
%
Case~1) If $1\geq \beta$: $X_2=X_0=0$, 
i.e., the information bits for destination~1 are carried by $X_1$.
The achievable rate is $r_1=1,r_2=0$.
Case~2) If $1< \beta$: $X_2=X_1=0$, 
i.e., the information bits for destination~1 are carried by $X_0$.
The achievable rate is $r_1=\beta,r_2=0$.
The other corner point $(r_1,r_2)=(0,\max\{1,\beta\})$ is achieved by swapping the role of the users.
By time-sharing, 
the whole dominant face of the outer bound region is achievable, thus proving capacity.

\begin{rem}
The points $(r_1,r_2)=(\max\{1,\beta\},0)$ and $(r_1,r_2)=(0,\max\{1,\beta\})$ are always corner points of the capacity region, but are not the  dominant ones in general.
\end{rem}

\paragraph{Case $n_{\rm{S}}>0$, $\alpha\not=1$ and $\max\{\alpha,\beta\} > 1$ (Regimes I to IV in Fig. \ref{fig:Regions})}
When \eqref{eq:DT OK 12}, \eqref{eq:DT OK 21} and \eqref{eq:DT OK ETW} are redundant,
that is, for $\max\{\alpha,\beta\} > 1$ (all but Regimes~V and~VI in Fig. \ref{fig:Regions}),
the region in \eqref{eq:highsnrbounds:sym} simplifies to the pentagon region
\begin{align}
&r_1   \leq  \max\{1, \beta\}, \quad
 r_2   \leq  \max\{1, \beta\}, \quad   
 r_1+r_2  \leq  \beta+\max\{1, \alpha\}.
\label{eq:all but regime V and VI}
\end{align}
We show achievability with two different strategies.

{\it b.1) Regimes II, III, and IV in Fig.~\ref{fig:Regions}.}
For $\beta\geq 1$, in order to prove capacity, it suffices to show the achievability of the corner point $(r_1,r_2) = (\beta,\min\{\beta,\max\{1,\alpha\}\})$. The other corner point $(r_1,r_2)=(\min\{\beta,\max\{1,\alpha\}\},\beta)$ is achieved by swapping the role of the users. By time-sharing between the corner points, the whole dominant face of the outer bound region is achievable, thus proving capacity.
%
Let $U_0,U_{1p},U_{2p}$ be independent vectors. 
Consider the following strategy
\begin{align}
      X_1 = \mathbf{S}^{m-n_{\rm{C}}} U_{1p},
\quad X_2 = \mathbf{S}^{m-n_{\rm{C}}} U_{2p},
\quad X_0 = \mathbf{S}^{m-n_{\rm{I}}} (U_{1p} + U_{2p})  +  U_0,
\label{eq:neutralize interference over the air}
\end{align}
where $X_0$ is so as to {\it  neutralize over the air} the interference at the destinations.
The received signal at destination~1 is
\begin{align}
 Y_1 &=
\underbrace{\Big(\mathbf{S}^{m-n_{\rm{S}}} \mathbf{S}^{m-n_{\rm{C}}}
   + \mathbf{S}^{m-n_{\rm{C}}} \mathbf{S}^{m-n_{\rm{I}}} \Big)}_{\not=\mathbf{0} \ \text{if} \ n_{\rm{S}}\not=n_{\rm{I}} \ \Longleftrightarrow \ \alpha\not=1 } U_{1p}
+\underbrace{\Big(\mathbf{S}^{m-n_{\rm{C}}} \mathbf{S}^{m-n_{\rm{I}}}
 + \mathbf{S}^{m-n_{\rm{I}}} \mathbf{S}^{m-n_{\rm{C}}} \Big)}_{=\mathbf{0}, \ \text{interference neutralized}} U_{2p}
 + \mathbf{S}^{m-n_{\rm{C}}} U_0 \notag
   \\&= \mathbf{S}^{m-n_{\rm{C}}}\left(
\Big(\mathbf{S}^{m-n_{\rm{S}}}  + \mathbf{S}^{m-n_{\rm{I}}} \Big) U_{1p} + U_0
\right), \label{eq:caseB}
\end{align}
and similarly for the received signal at destination~2.
Let the top $m-\max\{n_{\rm{S}},n_{\rm{I}}\}$ bits of $U_0$, which are received clean on top of the bits of $\Big(\mathbf{S}^{m-n_{\rm{S}}}  + \mathbf{S}^{m-n_{\rm{I}}} \Big) U_{ip}$ at each destination~$i\in[1:2]$, be i.i.d. Bernoulli($1/2$) bits dedicated to user~1 and the rest of $U_0$ be set to zero. $U_{1p}$ and $U_{2p}$ are i.i.d. Bernoulli($1/2$) bits. Hence, $R_1 = n_{\rm{C}}$ and $R_2 = \min\{n_{\rm{C}},\max\{n_{\rm{S}},n_{\rm{I}}\}\}$ (note that the rates cannot be larger than $n_{\rm{C}}$ because of the multiplication by $\mathbf{S}^{m-n_{\rm{C}}}$ of the signals at each receiver \eqref{eq:caseB}). By normalizing the rates by $n_{\rm{S}}$ the claim follows.

Notice that by setting $U_0=0$ and by using the ``neutralize over the air'' technique in~\eqref{eq:neutralize interference over the air}, it is always possible to achieve the normalized private rates $r_1=r_2=\min\{\beta,\max\{1,\alpha\}\}$, where we use the qualifier ``private'' to follow the nomenclature convention for the classical IC: a message that is decoded  only at an intended destination is referred to as a``private message.'' A message also decoded at a non intended destination is referred to as a ``common message.''   In Regimes II to IV, a ``common message'' for user~1 is sent by the CR through the top bits of $U_0$ whenever $\beta > \max\{1,\alpha\}$.

\begin{rem}
The achievability in this case can also be used to show the achievability of the region in~\eqref{eq:dof ns=0} when $n_{\rm{S}}=0$, in which case
the region in \eqref{eq:highsnrbounds:sym} simplifies to (here we do not normalize by the strength of the direct link as this link does not exist)
\begin{align}
R_1  \leq  n_{\rm{C}},  \quad 
R_2  \leq  n_{\rm{C}},  \quad 
R_1+R_2 \leq  n_{\rm{C}}+n_{\rm{I}}.
\label{eq:dof ns=0}
\end{align}
Clearly this is a special case of $1 < \max\{\alpha,\beta\}$ since $0=n_{\rm{S}} \leq \max\{n_{\rm{I}},n_{\rm{C}}\}$.
\end{rem}

{\it b.2) Regime I in Fig.~\ref{fig:Regions}.}
For $\beta < 1$ (and as a consequence of $\max\{\alpha,\beta\}>1$ we must have $\alpha>1$), in order to prove capacity, it suffices to show the achievability of $(r_1,r_2) = (1,\min\{1,\beta+\alpha-1\})$. 
Here we build on the observation made for the achievable scheme in Regimes II to IV and develop a scheme that in addition to the ``private rates'' $r_{1p}=r_{2p}=\min\{\beta,\max\{1,\alpha\}\}=\beta$ also conveys common rates $r_{1c} = 1-\beta$ and $r_{2c} = \min\{\alpha-1,1-\beta\}$.
In this regime some interfering bits can be decoded because the interference is strong ($\alpha>1$) at the non-intended destination. As opposed to Regimes II to IV where the ``common bits'' were carried by the CR though $U_0$, here they will be carried by $X_1$ and $X_2$, i.e., cooperation through the CR in this regime is too weak and it is better used to neutralize the interference rather than to deliver common bits.
Let $U_{1c},U_{1p},U_{2c},U_{2p}$ be independent vectors. 
Consider 
\begin{align}
 X_1  = \mathbf{S}^{m-n_{\rm{C}}} U_{1p} +  U_{1c}, \quad
 X_2  = \mathbf{S}^{m-n_{\rm{C}}} U_{2p} +  U_{2c}, \quad
 X_0  = \mathbf{S}^{m-n_{\rm{I}}} (U_{1p} + U_{2p}).
\end{align}
The received signal at destination~1 is
\begin{align}
 Y_1 &=
     \mathbf{S}^{m-n_{\rm{S}}} U_{1c}
   + \mathbf{S}^{m-n_{\rm{I}}} U_{2c} 
   + \mathbf{S}^{m-n_{\rm{C}}} \Big(\mathbf{S}^{m-n_{\rm{S}}}
   + \mathbf{S}^{m-n_{\rm{I}}} \Big) U_{1p},
\end{align}
and similarly for 
destination~2.
Clearly, 
if only the top $n_{\rm{S}}(1-\beta)$ bits of $U_{1c}$ are non-zero
and the top $n_{\rm{S}}\min\{\alpha-1,1-\beta\}$ bits of $U_{2c}$ are non-zero, then
destination~1 can decode $U_{2c},U_{1c},U_{1p}$ in this order 
and destination~2 can decode  $U_{1c},U_{2c},U_{1p}$ in this order, thus achieving the desired rates.

\begin{rem}\label{rem:wsibothlda}
Interestingly, the region in~\eqref{eq:all but regime V and VI} is equivalent to the capacity region under ``strong interference at both receivers'' in \cite[Theorem~V.2]{riniIFCCR}, 
defined as the channel parameters for which
\begin{align}
I(Y_2 ;      X_2, X_c |  X_1 ) \leq I(Y_1 ;      X_2 , X_0 | X_1 ), \quad 
I(Y_1 ;      X_1, X_c |  X_2 ) \leq I(Y_2 ;      X_1 , X_0 | X_2 ), 
\label{StrongInt:Conditions:Gaussian}
\end{align}
hold for all distributions that factor as $\mathbb{P}_{X_1,X_2,X_0}=\mathbb{P}_{X_1}\mathbb{P}_{X_2}\mathbb{P}_{X_0|X_1,X_2}$.
Evaluation of the condition of ``strong interference at both receivers''  in \eqref{StrongInt:Conditions:Gaussian} is difficult because all possible input distributions must be tested---or an argument must be found that allows  restriction to a specific subset of input distributions without loss of generality. For the LDA, it was not clear that i.i.d. Bernoulli($1/2$) input bits at all nodes would exhaust all possible input distributions, as this does not capture the possible correlation between $X_0$ and $(X_1,X_2)$.
It is interesting to notice that, with i.i.d. Bernoulli($1/2$) input bits at all terminals, that  the condition of ``strong interference at both receivers''  in \eqref{StrongInt:Conditions:Gaussian} gives $\max\{\alpha,\beta\} \geq \max\{1,\beta\}$, or equivalently,  $\max\{\alpha,\beta\} \geq 1$.
\end{rem}

\paragraph{Case $n_{\rm{S}}>0$, $\alpha\not=1$ and $\max\{\alpha,\beta\} \leq 1$: sub-case $0\leq\alpha\leq\beta\leq1$ (Regime V in Fig. \ref{fig:Regions})}
In Regime~V
the outer bound region is a square and has only one dominant corner point given by $r_1=r_2=1$.
Let $U_{1p},U_{2p}$ be independent vectors and set
\begin{align}
 X_1 = U_{1p}, \quad X_2 = U_{2p}, \quad
 X_0 = \mathbf{S}^{n_{\rm{C}}-n_{\rm{I}}} (U_{1p} + U_{2p}), 
\label{eq:sig regime V}
\end{align}
so as to neutralize  the interference at the destinations (note the different shifts of the ``private codewords'' as compared to the scheme for Regimes I to IV in Fig. \ref{fig:Regions}). In this regime $m = \max\{n_{\rm{S}},n_{\rm{C}},n_{\rm{I}}\}=n_{\rm{S}}$.
The received signal at destination~1 is
\begin{align}
 Y_1 &= \Big( \mathbf{S}^{m-n_{\rm{S}}} + \mathbf{S}^{m-n_{\rm{C}}} \mathbf{S}^{n_{\rm{C}}-n_{\rm{I}}} \Big) U_{1p},
\end{align}
and similarly for the received signal at destination~2.
Hence 
$R_1=R_2=n_{\rm{S}}\max\{1,\alpha\}=n_{\rm{S}} \cdot 1$.
By normalizing the rates by $n_{\rm{S}}$ the claim follows.

\begin{rem}
Notice that $X_0$ in~\eqref{eq:sig regime V} is obtained by downshifting $U_{1p} + U_{2p}$ by $n_{\rm{C}}-n_{\rm{I}}$ positions, or in other words, the top $n_{\rm{C}}-n_{\rm{I}}$ bits of $X_0$ are zero.
This strategy is slightly counter-intuitive as the cognitive relay, with knowledge of all messages, should be able to use all its bits without harm.  However, including bits here would not improve rates as the direct link is already able to convey these bits directly, and the cognitive relay is only really needed to simultaneously cancel the interference at both receivers. The desired signal can be obtained by multiplying the received signal by the inverse of $\mathbf{S}^{m-n_{\rm{S}}}+\mathbf{S}^{m-n_{\rm{I}}}$, which is well defined as long as $n_{\rm{S}}\not=n_{\rm{I}} \ \Longleftrightarrow \ \alpha\not=1$. 
\end{rem}

\paragraph{Case $n_{\rm{S}}>0$, $\alpha\not=1$ and $\max\{\alpha,\beta\} \leq 1$: sub-case $0\leq\beta\leq\alpha<1$ (Regime VI in Fig. \ref{fig:Regions})}
In Region~VI in Fig. \ref{fig:Regions},
the region in \eqref{eq:highsnrbounds:sym} simplifies to
\begin{subequations}
\begin{align}
r_1     &\leq 1, \quad
r_2     \leq 1,\\
r_1+r_2 &\leq 2-\alpha +\beta, \label{eq:regime VI sum1}\\
r_1+r_2 &\leq  2\max\{1-\alpha,\alpha\}+ 2\beta, \label{eq:regime VI sum2}\\
2r_1+r_2&\leq  1 + \max\{1-\alpha,\alpha\}
+ \max\{1-\alpha,\beta\} + 2\beta, \label{eq:regime VI sum21}\\
r_1+2r_2&\leq  1 + \max\{1-\alpha,\alpha\}
+ \max\{1-\alpha,\beta\} + 2\beta. \label{eq:regime VI sum12}
\end{align}
\label{eq:regime VI}
\end{subequations}
Due to the complexity of the outer bound region in \eqref{eq:regime VI},
Regime VI is further divided  into four sub-regimes, which also correspond to a generalization of the division of the W-curve in \cite{etkin_tse_wang} as $\beta$ is relatively small in this regime.
The boundary
     between Regimes VI.1 and VI.2 occurs at $       2\alpha=1$,
that between Regimes VI.2 and VI.3        at $\beta+ 3\alpha=2$, and
that between Regimes VI.3 and VI.4        at $\beta+  \alpha=1$.
These boundaries reduce to those of the W-curve in weak interference for $\beta=0$.
So far we were unable to show capacity for the whole Regime VI.
We propose next a capacity achieving scheme 
Regime VI.1 and discuss strategies for the remaining cases.

\begin{figure}
\centering
        \begin{subfigure}[b]{0.420\textwidth}
                \includegraphics[width=\textwidth]{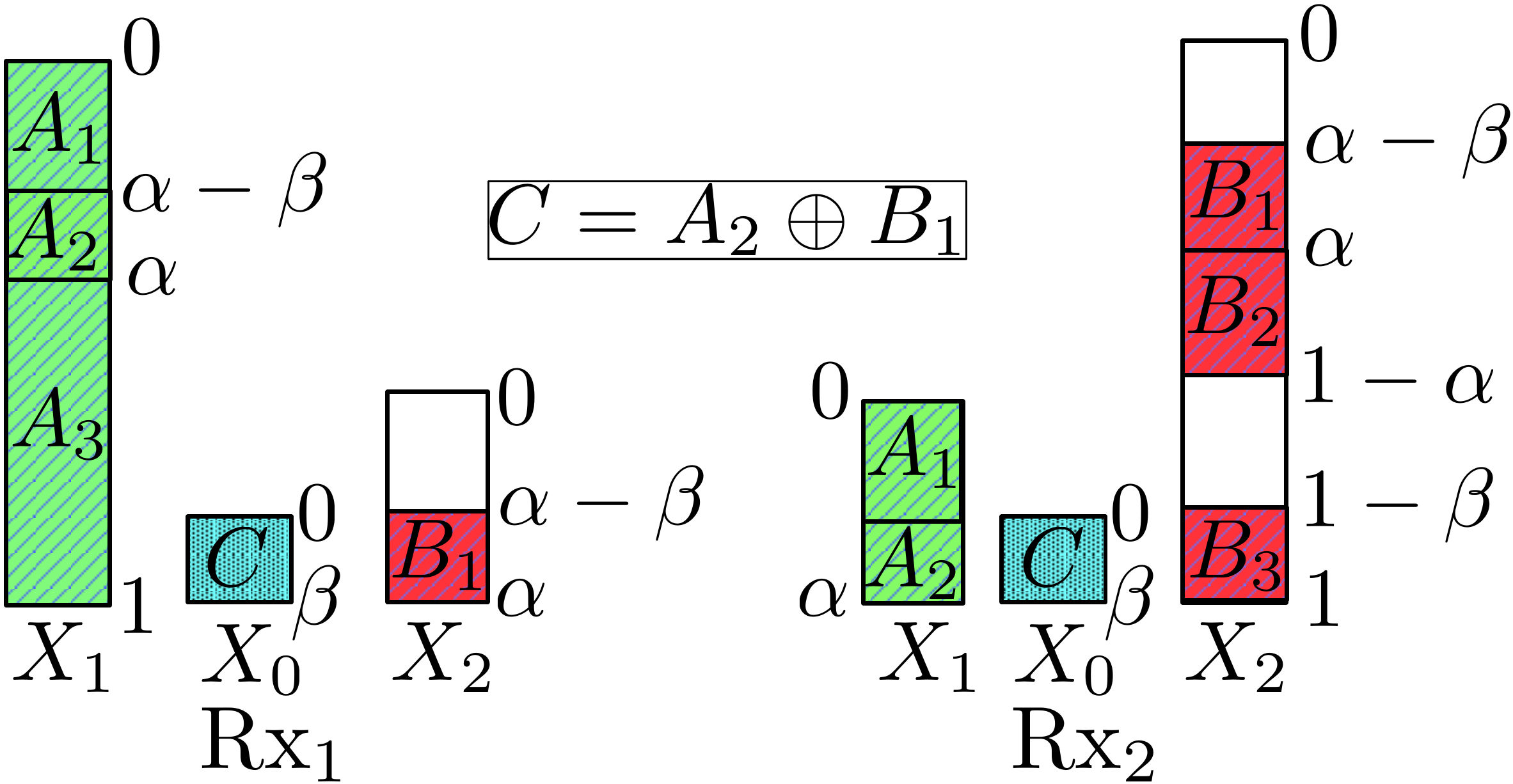}
                \caption{\small Optimal Strategy for Regime VI.I.}
                \label{fig:cap:and:strategyVI.1}
        \end{subfigure}
\hspace*{0.55cm}
        \begin{subfigure}[b]{0.520\textwidth}
                \includegraphics[width=\textwidth]{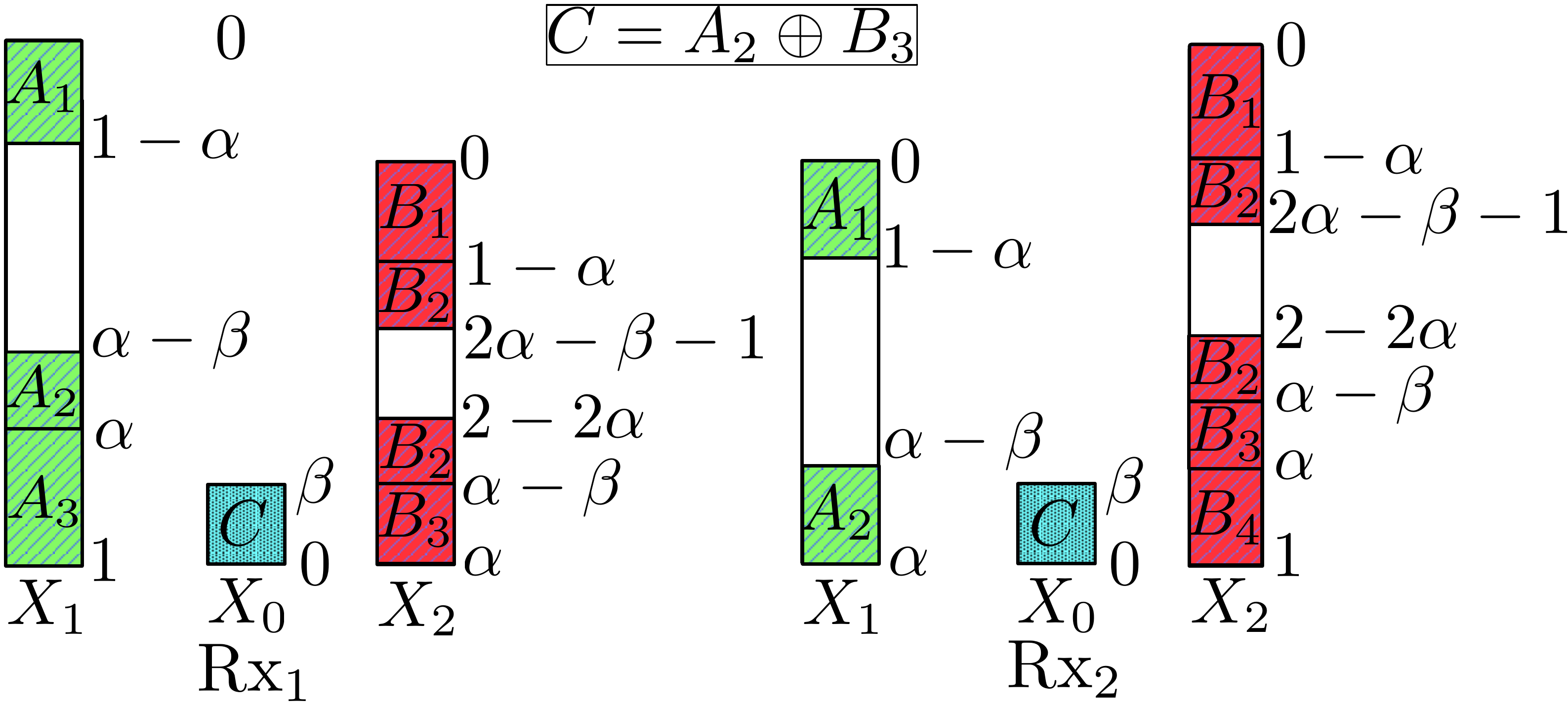}
                \caption{\small Sum-capacity optimal strategy for parts of Regimes VI.3 and VI.4.}
                \label{fig:SumCapVI}
        \end{subfigure}
        %
\caption{\small Achievability Strategies for portions of Regime VI of Fig.~\ref{fig:Regions}. }
\label{fig:alltogether for though regime}
\end{figure}

In Regime VI.1 ($0\leq \beta \leq \alpha \leq \frac{1}{2}$) capacity can be proved by showing the achievability of the corner point $(r_1,r_2)=(1, \ 1-2\alpha+2\beta)$ from \eqref{eq:regime VI}, {because in this regime the bounds in~\eqref{eq:regime VI sum1},~\eqref{eq:regime VI sum21} and~\eqref{eq:regime VI sum12} are redundant.} We will demonstrate our achievable scheme by using the graphical representation proposed in \cite{bresler_tse}. 
Fig.~\ref{fig:cap:and:strategyVI.1} shows such a strategy.
The blocks represent the signals arriving at each destination, where block lengths has been normalized by $n_{\rm{S}}$.  
Due to the channel downshift operation,
the desired signal at a destination has normalized length of $1$, the signal from a cognitive relay  has normalized length $\beta$, and the interfering signal  has normalized length $\alpha$.  Bits intended for destination~1 are denoted by $A_i$, and those destined to destination~2 by $B_i$, $i\in[1:3]$.
In our example, the relay sends $C := A_2\oplus B_1$, where $A_2$ and $B_1$ play the role of $U_{1p}$ and $U_{2p}$, respectively, in the previous regimes, i.e., they are ``private bits'' whose effect is ``neutralized over the air'' by the relay. Source~1 sends the block of bits indicated as $A_1$ ``on top'' of $A_2$; here $A_1$ plays role of $U_{1c}$ in the previous regimes, i.e., they are ``common bits'' decodes at both destination; as for the classical IC, bits $A_2$ can be decoded at destination~2 if they are received interference-free at destination~2~\cite{bresler_tse}, which is possible thanks to the fact that a portion of the signal sent by source~2 contains zeros (in between blocks $B_2$ and $B_3$). Blocks $A_3,B_2$ and $B_3$ are ``private bits'' too. However these bits do not require ``neutralization'' by the relay as they appear ``below the noise floor'' at the non-intended receiver (similarly to the classical IC~\cite{bresler_tse}, these bits are actually not received at the non-intended destination). Notice that the top portion of the signal sent by source~2 is also populated by zeros (above block $B_1$); this is needed to allow destination~1 to decode $A_3$.
%
Destination/Rx1 decodes $A_1,A_2,A_3$ in this order, as does not suffers any interference from user~2, and achieves normalized rate $r_1=1$.
Destination/Rx2 decodes $B_1,B_2,A_1,B_3$ in this order, and achieves normalized rate $r_2=1-2\alpha+2\beta$.
%
%

\paragraph{On Capacity and sum-capacity for parts of Regimes VI.3 and VI.4 in Fig.~\ref{fig:Regions}}
Fig.~\ref{fig:SumCapVI} shows an achievable scheme for the case
$4\alpha -3 \leq \beta  \le 3\alpha -2, \  2/3 \leq \alpha \leq 1$,
where the restriction of the possible values of $(\alpha,\beta)$ is due to the fact that certain pieces of $X_2$ must have non-negative length. 
The corner point we aim to achieve is  $(r_1,r_2)=(1, 1-\alpha+\beta)$.
Because the outer bound region in Regime VI.4, described by
\begin{subequations}
\begin{align}
&r_1  \leq  1, \quad
 r_2  \leq  1, \quad
r_1+r_2  \leq  2-\alpha +\beta,
\end{align}
\label{cap:Regime VI.4}
\end{subequations}
has only two corner points, achieving one of them implies the achievability of the entire capacity region by a time sharing argument.
In contrast,  Regime VI.3 described by
\begin{subequations}
\begin{align}
&r_1  \leq  1, \, r_2  \leq 1, \label{eq:sumCapacitysinglerates}\\
&r_1+r_2  \leq  2-\alpha +\beta, \label{eq:sumCapacity}\\
&2r_1+r_2 \leq  1 + \alpha +\max\{1-\alpha,\beta\}+2\beta, \label{eq:CornerPoints}\\
&r_1+2r_2 \leq  1 + \alpha + \max\{1-\alpha,\beta\}+2\beta,
\end{align}
\label{cap:Regime VI.3}
\end{subequations}
has four corner points and thus achieving $(r_1,r_2)=(1, 1-\alpha+\beta)$ does not suffice to show capacity.
The other dominant corner point in Regime VI.3 is determined by the intersection of the $r_1$-bound in~\eqref{eq:sumCapacitysinglerates} with the $(2r_1+r_2)$-bound \eqref{eq:CornerPoints}.
Thus, the strategy in Fig.~\ref{fig:SumCapVI} is only sum-rate optimal and  works in the following way.
Destination/Rx1 decodes the desired vector $A_1$ and the undesired vectors $B_1$ and $B_2$.
Now, since $B_2$ has been decoded, it can be subtracted at the points where its repetition interference with $A_2$ and $A_3$.
Thus, $A_2$ and $A_3$ are decoded too (note that the effect of $B_3$ has been ``neutralized'' by the relay and the block $A_2$ is decoded before decoding $A_3$ so its effect can be removed from $A_3$).
%
Destination/Rx2 first decodes $B_1$. Next, because vectors $B_2$, $B_3$  and $B_4$ do not experience any interference, they can be decoded as well.
Finally, since $B_2$ has been already decoded, the portion where $A_1$ interferes with $B_2$ can be ignored.
This achieves $r_1+r_2  =  2-\alpha +\beta$.
\end{IEEEproof}

It would be interesting to know what could be missing for Regimes VI.2 to VI.4.
The capacity region for Regimes VI.2 to VI.4 remains unknown. These regimes are related to the most involved region of the W-curve in  \cite{etkin_tse_wang} for the IC in moderately weak interference (i.e., for $\alpha\in[1/2,1]$) and as such it is not surprising that these are also the most difficult cases for the ICCR.
At this point we conjecture that
the way we have bounded the multi-letter portion (MLP) in~\eqref{eq:mlp:singleletter:attempt2}
is too loose.
We note that the entropy of a discrete random variable is non-negative and is not decreased by removing conditioning.
Possibly the bound in \eqref{eq:mlp:singleletter:attempt2} does not accurately capture the correlation between $X_0$ and $(X_1,X_2)$. Essentially the bound in \eqref{eq:mlp:singleletter:attempt2}, which for the symmetric LDA is given in~\eqref{eq:outer bound:LD:mlp}, appears to assert that $X_0$ can be simultaneously maximally correlated with both $X_1$ and $X_2$.
However, if $X_0$ is maximally correlated with $X_1$, i.e., $X_0=X_1$, then it is independent of $X_2$
(recall that $X_1$ and $X_2$ are independent because carry independent messages); in this case
the $\mathsf{MLP}$ expression would be $\min\{\alpha,\beta\}$ rather than $2 \min\{\alpha,\beta\}$.
Tightening the bounds in \eqref{eq:DT OK ETW}, \eqref{eq:DT OK 12} and \eqref{eq:DT OK 21}
so as to capture the correlation among transmitted signals,
and/or to derive another bound of the form $2R_1+R_2$ or $R_1+2R_2$ (such a bound was needed for the IC
with rate-limited receiver cooperation \cite{wang2010interference}) is the subject of ongoing investigation. 


\section{Approximate Capacity for the symmetric GICCR}
\label{sec:Gaussian}

We now concentrate our attention on the symmetric GICCR.
We will use the insights gained from the symmetric LDA to prove a constant gap result in those regimes where capacity is not known~\cite{riniIFCCR}. Tthe symmetric GICCR is parameterize as
\begin{subequations}
\begin{align}
   |h_{11}|^2=|h_{22}|^2=|h_{\rm{S}}|^2&:=\mathsf{SNR}^1,
\\ |h_{12}|^2=|h_{12}|^2=|h_{\rm{I}}|^2&:=\mathsf{SNR}^\alpha, \ \alpha \geq 0,
\\ |h_{20}|^2=|h_{10}|^2=|h_{\rm{C}}|^2&:=\mathsf{SNR}^\beta,  \ \beta \geq 0.
\end{align}
\label{eq:GICCR snr exponents interf}
\end{subequations}
where here $\alpha$ and $\beta$ have  meaning similar to the parameters used in the LDA model in \eqref{eq:parameter LDA}, in particular
$\alpha$ is the ratio of the received power on the interference link expressed  in dB over the received power on the direct link expressed  in dB, and $\beta$ is the ratio of the received power on the relay-destination link expressed in dB over the received power on the direct link expressed in dB. The normalization of the SNR-exponent of the direct link to $1$ is without loss of generality and parallels the normalization by $n_{\rm{S}}$ in the LDA. The following results parallel Theorem~\ref{thm:LDA:cap}:

\subsection{Capacity in Regimes I to IV in Fig.~\ref{fig:Regions}}
Recently, the capacity of~\eqref{StrongInt:Conditions:Gaussian} was characterized in the ``strong interference at both receivers'' \cite{riniIFCCR}, which in the symmetric GICCR reduces to \cite[eq.(27)]{riniIFCCR}
\footnote{The detailed proof is rather involved and uses the so called extremal inequality \cite{EPIjournal}. The main difficulty arises from the fact that $X_1$ and $X_2$ are correlated with $X_0$ and a more elaborate argument to show that Gaussians are optimal is needed. For interested readers the proof may be found in \cite[Theorem VI.1]{riniIFCCR} .}
\begin{align}
&\Big| |h_{\rm{S}}|+|h_{\rm{C}}| \Big|^2 \le \Big| |h_{\rm{I}}|{\rm e}^{+{\rm j} \theta}+|h_{\rm{C}}| \Big|^2,
\quad \theta \in\{ \angle{h_{12}}, \angle{h_{21}}\},
\label{StrongInt:Conditions:Gaussian:Symmetric}
\end{align}
where $\angle{h_{21}},\angle{h_{12}}$ are the phases of the cross-link channel gains (the ones that could not be taken to be real-valued and non-negative without loss of generality in~\eqref{eq:GICCR}).

Using \eqref{eq:GICCR snr exponents interf} and taking $\mathsf{SNR} \to \infty $ the condition in~\eqref{StrongInt:Conditions:Gaussian:Symmetric}, by assuming that $|h_{\rm{I}}|{\rm e}^{+{\rm j} \angle{h_{ij}}}+|h_{\rm{C}}|\not=0$, reduces to
\begin{align}
&\max\{\mathsf{SNR},\mathsf{SNR}^{\beta}\} \le \max\{\mathsf{SNR}^{\alpha},\mathsf{SNR}^{\beta}\}
\Longleftrightarrow 1 \le \max\{\alpha,\beta\}.
\label{eq:highSNR:Regimes}
\end{align}
The high-SNR regime of~\eqref{eq:highSNR:Regimes} coincides with Regimes I to IV in Fig.~\ref{fig:Regions} for the LDA (see also Remark~\ref{rem:wsibothlda}). In \cite{riniIFCCR} it was shown that joint decoding of all messages at both receivers is optimal or capacity achieving when the ``strong interference at both receivers'' condition is satisfied.
We therefore concentrate here on mimicking, in the Gaussian case, those regimes for which we could prove capacity in the LDA, namely Regime V and VI.1.

\subsection{Capacity to Within a Constant Gap in Regime V in Fig.~\ref{fig:Regions}}
Regime~V in the LDA is characterized by $\alpha \le \beta \le 1$, which we try to match with something of the form $|h_{\rm{I}}|^2 \le |h_{\rm{C}}|^2 \le |h_{\rm{S}}|^2$ for the GICCR. We now  build on the intuition developed for the LDA and propose a simple scheme that is optimal to within an additive gap. 
\begin{thm}
\label{thm:constanGap}
For the symmetric GICCR, the capacity outer bound in Theorem~\ref{thm:outer bound:DM} is achievable to within $\log_2\left({4}/{(1-\frac{1}{\sqrt{2}})^2}\right)\approx 5.5$~bits per user for $2|h_{\rm{I}}|^2 \le |h_{\rm{C}}|^2 \le |h_{\rm{S}}|^2$.
\hfill$\square$\end{thm}
\begin{IEEEproof}
In Regime V for the LDA, the cognitive relay simultaneously neutralizes over the air the interference at both receivers.
This mode of operation is reminiscent of zero forcing. We therefore propose: let $U_{1p}$ and $U_{2p}$ be two independent Gaussian random variables with zero mean and unit variance and define for some $(\rho_1,\rho_2)$ such that $|\rho_1|^2+|\rho_2|^2 \le 1$
\begin{align}
X_1=U_{1p}, \quad
X_2=U_{2p}, \quad
X_0=\rho_1 U_{1p}+\rho_2 U_{2p}.
\end{align}
Next we choose $\rho_1$ and $\rho_2$ so as to simultaneously neutralize the contribution of $U_{2p}$ at destination~1 and of $U_{1p}$ at destination~2. This is possible if
\begin{align}
 \rho_1 = -\frac{|h_{\rm{I}}|{\rm e}^{+{\rm j} \angle{h_{21}}}}{|h_{\rm{C}}|}, \quad
 \rho_2 = -\frac{|h_{\rm{I}}|{\rm e}^{+{\rm j} \angle{h_{12}}}}{|h_{\rm{C}}|}, \quad
\label{eq:condForZF}
\end{align}
which requires $2|h_{\rm{I}}|^2 \le |h_{\rm{C}}|^2$.
With this assignment the channel outputs become
\begin{align}
&Y_1
=\big(|h_{\rm{S}}|-|h_{\rm{I}}|{\rm e}^{+{\rm j} \angle{h_{21}}}\big)U_{1p}+Z_1,
\quad 
Y_2
=\big(|h_{\rm{S}}|-|h_{\rm{I}}|{\rm e}^{+{\rm j} \angle{h_{12}}}\big)U_{2p}+Z_2,
\end{align}
%
and thus the following rates are achievable
\begin{align}
R_1 \le \log\left(1+\big||h_{\rm{S}}|-|h_{\rm{I}}|{\rm e}^{+{\rm j} \angle{h_{21}}}\big|^2\right),\quad
R_2 \le \log\left(1+\big||h_{\rm{S}}|-|h_{\rm{I}}|{\rm e}^{+{\rm j} \angle{h_{12}}}\big|^2\right).
\label{eq:ZF:innerbound}
\end{align}
From the outer bound we have
%
%
\begin{align}
R_1 &\leq I(Y_1;X_1,X_0|Q,X_2)
\leq \log\left( 1+\big(|h_{\rm{S}}|+|h_{\rm{C}}|\big)^2 \right)
\leq \log\left( 1+4\max\{|h_{\rm{S}}|^2,|h_{\rm{C}}|^2\} \right),
\label{eq:piecewise:outerbound}
\end{align}
and similarly for $R_2$.
Next, the argument of the log-function in~\eqref{eq:ZF:innerbound} can be lower bounded by
$(|h_{\rm{S}}| - |h_{\rm{I}}|)^2$.
Imposing  $|h_{\rm{C}}|^2 \le |h_{\rm{S}}|^2$, in order to mimic Regime~V of the LDA,  and $2|h_{\rm{I}}|^2 \le |h_{\rm{C}}|^2$, implies $|h_{\rm{I}}|^2 \le |h_{\rm{S}}|^2/2$, so that
$(|h_{\rm{S}}| - |h_{\rm{I}}|)^2 \geq \left(1-\frac{1}{\sqrt{2}}\right)^2 |h_{\rm{S}}|^2$.
Finally, by taking the difference between the upper bound in \eqref{eq:piecewise:outerbound}
and the lower bound in \eqref{eq:ZF:innerbound} we arrive at the claimed gap result.
\end{IEEEproof}

It is pleasing to see that a simple interference management technique reminiscent of zero-forcing is optimal to within a constant gap for the GICCR.  Notice that in this regime the channel behaves effectively as two non-interfering point-to-point links.

{

\subsection{Capacity to Within a Constant Gap in Regime VI.1 in Fig.~\ref{fig:Regions}}
\label{sec:VI.1gap}

Regime VI.1 for the LDA is characterized by $\beta \le \alpha \le \frac{1}{2}$, which we try to match with something of the form $|h_{\rm{C}}|^2\leq |h_{\rm{I}}|^2\leq \sqrt{|h_{\rm{S}}|^2}$ for the GICCR. Next, we build on the intuition developed in the LDA and propose a scheme that is optimal to within an additive gap.
\begin{thm}\label{thm:an achievable region in Regime VI.1}
For the symmetric GICCR, the capacity outer bound in Theorem~\ref{thm:outer bound:Gaussian} is achievable to within $\gapVIone$~bits per user if the channel gains satisfy the following three conditions:
(c1) $|h_{\rm{C}}|^2\leq |h_{\rm{I}}|^2\frac{|h_{\rm{I}}|^2}{1+|h_{\rm{I}}|^2}$,
(c2) $|h_{\rm{C}}|^2\leq \frac{1}{2}\frac{1+|h_{\rm{S}}|^2}{1+|h_{\rm{I}}|^2}$,
(c3) $|h_{\rm{I}}|^2(1+|h_{\rm{I}}|^2) \leq |h_{\rm{S}}|^2$,
(c4) $1\leq \min\{|h_{\rm{S}}|^2,|h_{\rm{I}}|^2\}$,
(c5) $\frac{|h_{\rm{S}}|^2}{1+|h_{\rm{I}}|^2}\geq 9$.
\hfill$\square$\end{thm}
\begin{IEEEproof}
The conditions (c1)-(c3) at high SNR are equivalent to $\beta \le \alpha \le \frac{1}{2}$;  conditions (c4)-(c5) are convenient for  gap computation.
In Regime VI.1 for the LDA, the CR simultaneously neutralizes interference at destination 1 and part of the interference at destination 2, see Fig.~\ref{fig:cap:and:strategyVI.1}.
We therefore propose the following choice of inputs: for $X_{A_1},X_{A_2},X_{A_3},X_{B_1},X_{B_2},X_{B_3}$ i.i.d. Gaussian random variables with zero mean and unit variance, let
\begin{align*}
X_1&=a_1X_{A_1}+a_2X_{A_2}+a_3X_{A_3}, & X_2&=b_1X_{B_1}+b_2X_{B_2}+b_3X_{B_3},& X_0&=\ell_1X_{A_2}+\ell_2X_{B_1},
\\
|a_1|^2&=\frac{|h_{\rm{I}}|^2}{1+|h_{\rm{I}}|^2}-\frac{|h_{\rm{C}}|^2}{2|h_{\rm{I}}|^2}  ,&
 b_1   &=\frac{|h_{\rm{C}}|}{\sqrt{2}|h_{\rm{I}}|{\rm e}^{+{\rm j} \angle{h_{12}}}},&
\ell_1 &=-\frac{1}{\sqrt{2}}, 
\\
 a_2   &=\frac{|h_{\rm{C}}|}{\sqrt{2}|h_{\rm{I}}|{\rm e}^{+{\rm j} \angle{h_{21}}}},&
|b_2|^2&=\frac{1}{1+|h_{\rm{I}}|^2}-\frac{|h_{\rm{C}}|^2}{1+|h_{\rm{S}}|^2},&
\ell_2 &=-\frac{1}{\sqrt{2}}, 
\\
|a_3|^2&=\frac{1}{1+|h_{\rm{I}}|^2},&
|b_3|^2&=\frac{|h_{\rm{C}}|^2}{1+|h_{\rm{S}}|^2}.
\end{align*}
Under the channel conditions
$|h_{\rm{C}}|^2 \leq 2|h_{\rm{I}}|^2 \frac{|h_{\rm{I}}|^2}{1+|h_{\rm{I}}|^2}$ so that $|a_1|^2\geq 0$, and
$|h_{\rm{C}}|^2 \leq \frac{1+|h_{\rm{S}}|^2}{1+|h_{\rm{I}}|^2}$ so that $|b_2|^2\geq 0$,
the transmitter power constraints are satisfied; 
these conditions are true by (c1) and (c2), respectively.
Note that transmitter~2 does not fully utilize its power.
%
With this choice of coefficients / power allocation, the channel outputs become
\begin{align*}
Y_1&=
 |h_{\rm{S}}|(a_1X_{A_1}+a_3X_{A_3})
+\frac{|h_{\rm{C}}|}{\sqrt{2}}\left(\frac{|h_{\rm{S}}|}{|h_{\rm{I}}|}{\rm e}^{-{\rm j} \angle{h_{21}}}-1\right)X_{A_2} 
+|h_{\rm{I}}|(b_2X_{B_2}+b_3X_{B_3}){\rm e}^{+{\rm j}\angle{h_{12}}}+Z_1,\\
Y_2&=
 |h_{\rm{I}}|(a_1X_{A_1}+a_3X_{A_3}){\rm e}^{+{\rm j} \angle{h_{21}}}
+\frac{|h_{\rm{C}}|}{\sqrt{2}}\left(\frac{|h_{\rm{S}}|}{|h_{\rm{I}}|}{\rm e}^{-{\rm j}\angle{h_{12}}}-1\right)X_{B_1} 
+|h_{\rm{S}}|(b_2X_{B_2}+b_3X_{B_3})+Z_2,
 \end{align*}
since $X_{B_1}$ has been zero forced at $Y_1$, and $X_{A_2}$ at $Y_2$, similarly to the scheme in Fig.~\ref{fig:cap:and:strategyVI.1} for the LDA.
By mimicking the corresponding scheme for the LDA, destination 1 successively decodes $X_{A_1},X_{A_2},X_{A_3}$ in this order, and destination 2 successively decodes $X_{B_1},X_{B_2},X_{A_1},X_{B_3}$ in this order; with this decoding procedure the following rates are achievable (see Appendix~\ref{app:VI.1gap})
\begin{align*}
R_{A_1}&= \log\left(1+\frac{|h_{\rm{I}}|^2}{4(3+|h_{\rm{C}}|^2)}\right),
&
R_{A_2}&= \log \left(1+\frac{|h_{\rm{C}}|^2}{10} \right),
&
R_{A_3}&= \log\left(1+\frac{|h_{\rm{S}}|^2}{1+2|h_{\rm{I}}|^2}\right),
\\
R_{B_1}&= \log \left(1+\frac{|h_{\rm{C}}|^2}{10} \right),
&
R_{B_2}&= \log \left(1+\frac{|h_{\rm{S}}|^2}{4(1+|h_{\rm{I}}|^2)^2} \right),
&
R_{B_3}&= \log\left(1+\frac{|h_{\rm{C}}|^2}{4}\right).
\end{align*}
We next compare this lower bound 
with the outer bound obtained by intersecting the sum-rate upper bound in~\eqref{eq:outer bound:inj sumrate} with the MLP tightened as in Theorem~\ref{thm:outer bound:Gaussian} (see Appendix~\ref{elavulationofETWsumrate}) and the single-rate upper bound in~\eqref{eq:out:DM1} (see eq.\eqref{eq:piecewise:outerbound}), that is, the corner point outer bound with coordinates
\begin{subequations}
\begin{align}
R_1 & =  \log \left( 1+4 |h_{\rm{S}}|^2 \right),\label{eq:ourterboud:CornerPont1}
\\
R_2 & = 2\log \left(
\left(1+|h_{\rm{I}}|^2+\frac{|h_{\rm{S}}|^2}{1+|h_{\rm{I}}|^2} \right)
\left(1+|h_{\rm{C}}|^2 \right) 2(1 + 1/\sqrt{2})^2
\right)
-\log \left( 1 +4 |h_{\rm{S}}|^2 \right). \label{eq:ourterboud:CornerPont2}
\end{align}
\label{eq:ourterboud:CornerPont2mainall}
\end{subequations}
In Appendix~\ref{app:VI.1gap} we show that the gap between
the inner and outer bound is at most $\gapVIone$~bits per user.
By swapping the role of the users, the other sum-capacity achieving corner point of the capacity region outer bound can be attained to within the same gap.

By setting $R_2=0$ and not using the CR we can achieve $R_1=\log(1+|h_{\rm{S}}|^2)$, which is at most 2~bits away from the corner point where $R_1$ is upper bounded by~\eqref{eq:ourterboud:CornerPont1} and $R_2=0$. The same reasoning holds with the role of the users swapped.
This shows that all corner points of the outer bound region can be achieved to within $\gapVIone$ bits per user. Therefore, by time sharing, the whole capacity region outer bound can be achieved to within a constant gap.
This concludes the proof.
\end{IEEEproof}

The gap in this regime is fairly large; we believe that this is due to the crude lower bounding steps for the achievable rates and to the simplicity of the proposed interference zero-forcing scheme. Numerical evaluations show that the actual gap when optimizing the power splits in the proposed scheme is actually lower.
}

\subsection{Numerical Comparisons} 
\begin{figure}
\centering
\includegraphics[width=10cm]{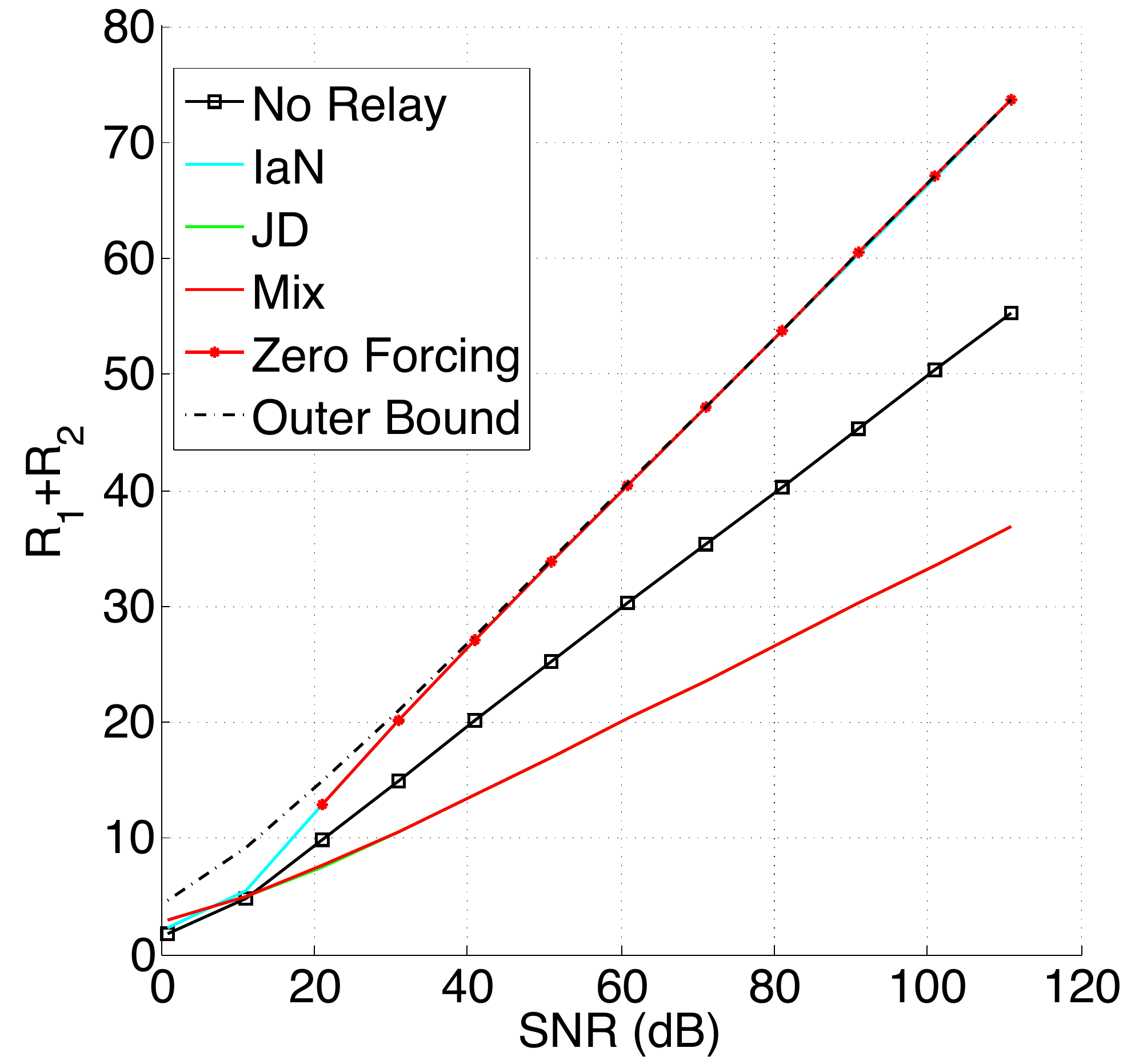}%
\caption{\small Numerical comparison of various  strategies for the GICCR for $\alpha=0.5, \beta=0.7$.
No relay: the relay is not used and the rates are given by the optimal interference channel strategy.
For the other curves the relay uses a linear strategy and the receivers apply
JD (jointly decode both messages), or
IaN (treat interference as noise), or
Mix (one receiver decodes both messages and the other only its intended one), or
ZF (relay performs zero forcing as in \eqref{eq:ZF:innerbound}, which is special case of IaN).
The outer bound is from~\eqref{eq:piecewise:outerbound}.}
\label{fig:compStrategy}
\end{figure}

\begin{figure}[h!]
\begin{subfigure}[t]{0.3\textwidth}
\centering
\includegraphics[width=8cm]{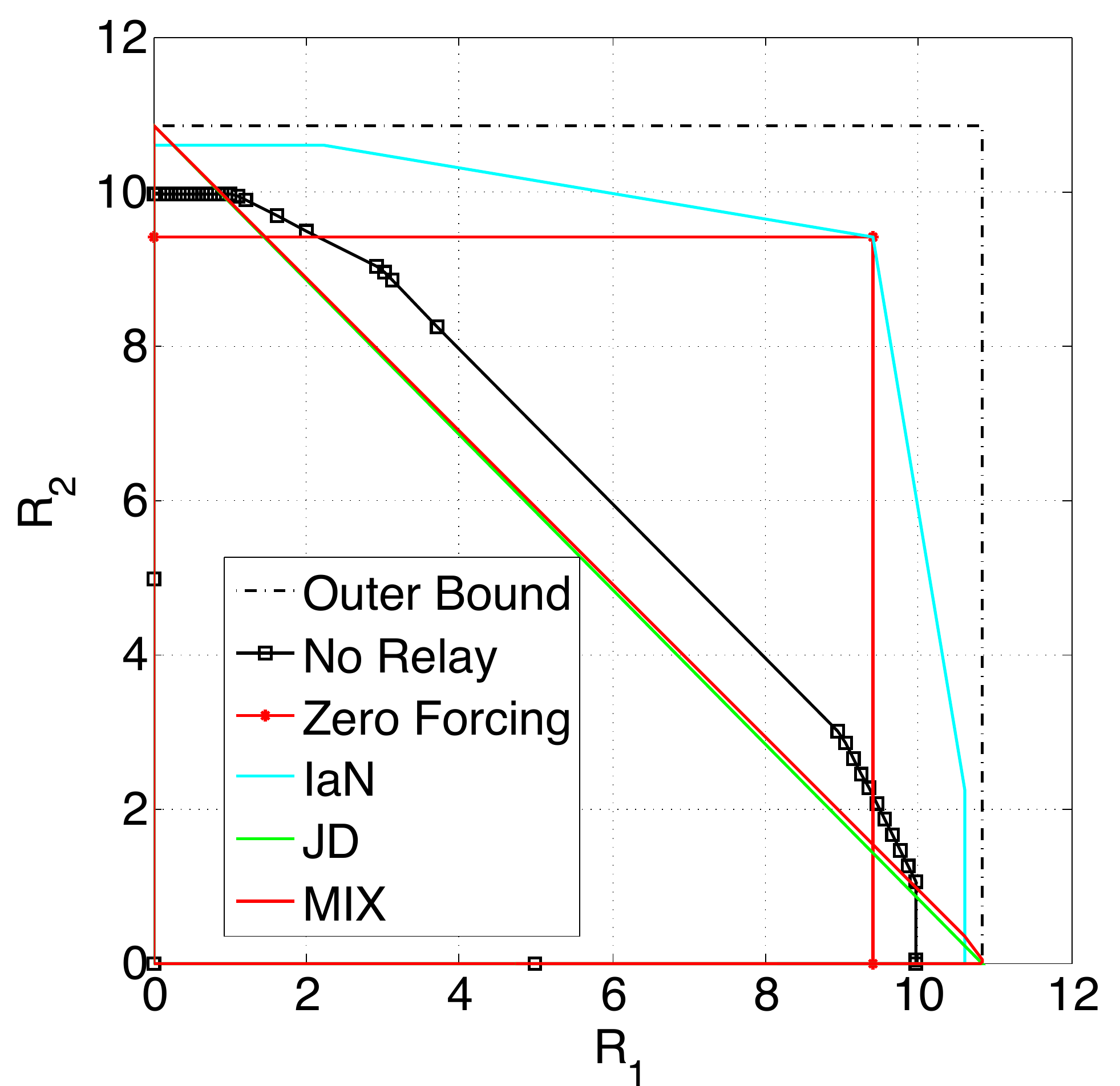}
\caption{$\mathsf{SNR}=30$dB.} 
\label{fig:Capa}
\end{subfigure}
\hspace*{4cm}
\begin{subfigure}[t]{0.3\textwidth}
\centering
\includegraphics[width=8cm]{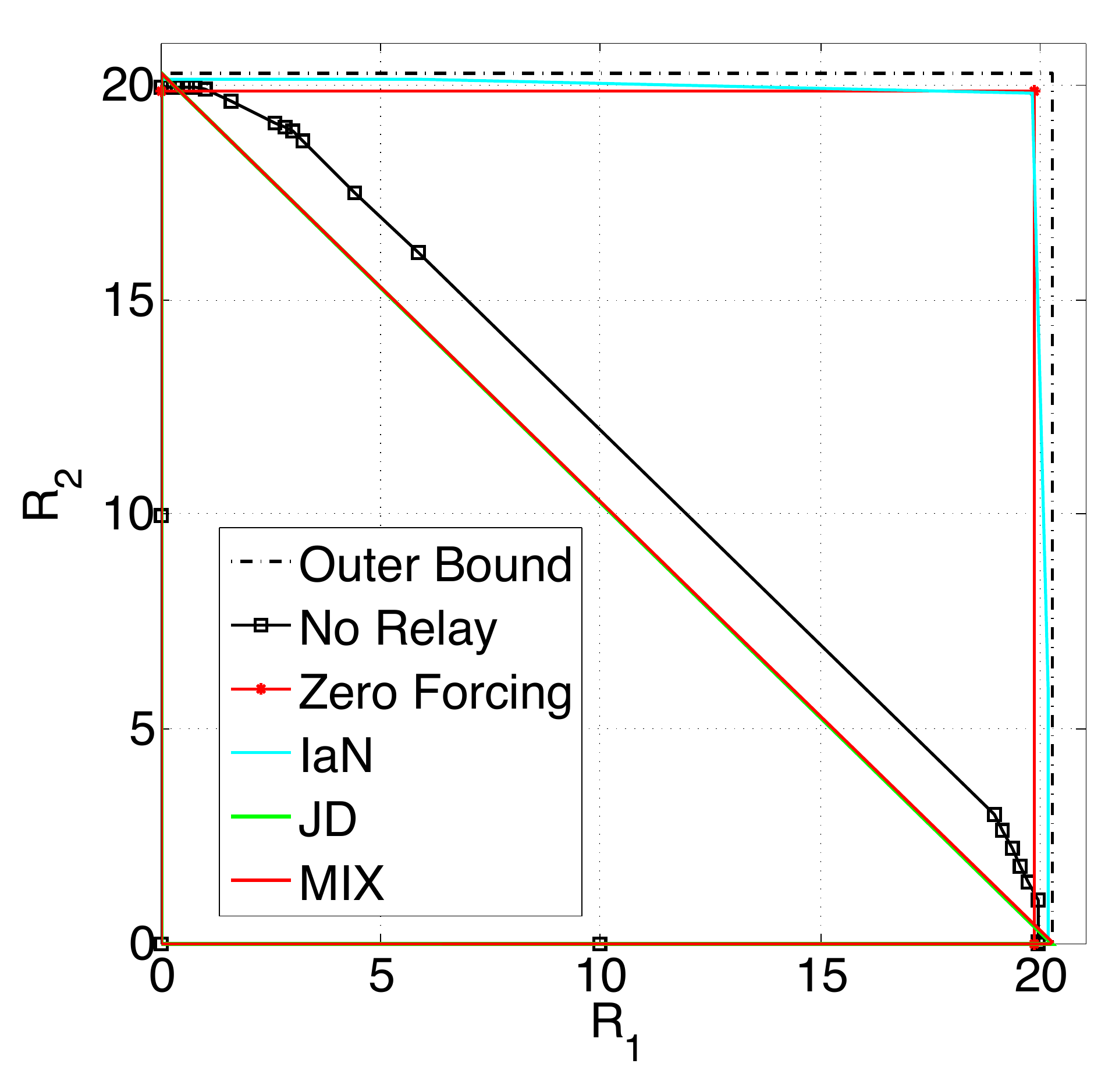}
\caption{$\mathsf{SNR}=60$dB.} 
\label{fig:CapacityRegionsNumerical}
\end{subfigure}

\caption{Achievable and outs bound regions for the GICCR with $\alpha=0.5, \beta=0.7$.
{ Same settings and conventions as in Fig.~\ref{fig:compStrategy}}. 
}
\label{fig:compStrategiesRegions}
\end{figure}

We conclude this section with some numerical examples.
Fig.~\ref{fig:compStrategy} and  Fig.~\ref{fig:compStrategiesRegions}  compare the performance of different achievable strategies as a function of the SNR (in dB) in Regime V, where the new constant gap result is obtained from Theorem~\ref{thm:constanGap}.
We note that the purpose of this paper is to provide {\it simple} achievable schemes for the Gaussian channel that are provably  optimal to within a constant gap, rather than focussing on finding the parameters that optimize the largest known (quite involved) achievable rate region for the ICCR derived in \cite [Theorem~IV.1]{riniIFCCR}. To this end, we compare several simple achievability schemes, including the constant gap to capacity scheme in~\eqref{eq:ZF:innerbound} and the  outer bound in~\eqref{eq:outer bound:DM}.

In Fig. \ref{fig:compStrategy}, we increase the $\mathsf{SNR}$ with fixed $\alpha=0.5$ and $\beta=0.7$ in~\eqref{eq:GICCR snr exponents interf} and compare the following strategies.
In the first strategy the relay stays silent and we use a well-known achievability strategy for the Gaussian IC (a version of the Han and Kobayashi strategy \cite{etkin_tse_wang}).
For the second achievability scheme, the relay is used and performs the simple linear combination scheme (rather than more complex schemes such as dirty paper coding) 
$X_0=a_1X_1+a_2X_2 : \  |a_1|^2+|a_2|^2 \le 1$,
where we optimize over $a_1$ and $a_2$.
We consider the following strategies at the receivers:
\begin{enumerate}
\item JD (Joint Decoding): both transmitters use common messages only, which are decoded at both destinations---the region thus looks like a compound multiple access channel with each message amplified at the receiver due to the relay's transmission.
\item IaN (Interference as Noise): destinations treat non-desired interference  as noise. All messages are therefore private.
\item Mix: one of the transmitters uses a common message and the other uses a private message; the common message is decoded at both receivers and the private is decoded at the appropriate receiver only and  treated as noise at the other.
\item ZF (Zero Forcing): use $a_1=-\frac{|h_{\rm{I}}|{\rm e}^{+{\rm j} \angle{h_{21}}}}{|h_{\rm{C}}|}$ and $a_2=-\frac{|h_{\rm{I}}|{\rm e}^{+{\rm j} \angle{h_{12}}}}{|h_{\rm{C}}|}$ as in Theorem~\ref{thm:constanGap} (when possible).
\end{enumerate}
Finally, the sum-rate outer bound from \eqref{eq:piecewise:outerbound} is plotted (i.e., in this case the whole capacity region is a square).

From  Fig.~\ref{fig:compStrategy} we see that as the SNR increases, the IaN and ZF schemes (ZF is actually one very specific choice of the IaN scheme where $a_1, a_2$ are specified explicitly) essentially overlap with the outer bound, which verifies the constant gap to capacity claim numerically. This scheme significantly outperforms (diverging slopes means the gap can be arbitrarily large) not using a relay at all, even with an optimizing transmission strategy, or using a JD or Mix strategy where the relay uses a simple linear combination scheme. 
Fig.~\ref{fig:compStrategiesRegions} shows the actual regions, rather than sum-rates, for the same settings and conventions as in Fig.~\ref{fig:compStrategy} for two different SNRs. 

We note that our goal is not to derive the best achievability scheme at any SNR, but rather to derive a simple, constant gap to capacity scheme and compare it to other, simple schemes.

\section{Conclusion}
\label{sec:conclusion}
We considered an interference channel in which a cognitive relay aids in the transmission of the two independent messages. We obtained the capacity region in almost all regimes for the symmetric LDA and translated these insights into a constant gap to capacity result for the corresponding Gaussian model. The capacity achieving schemes for the symmetric LDA use a variety of techniques at the cognitive relay, which  both aid in the transmission of the messages to the receivers, and  simultaneously neutralize interference at the two receivers.
Given the generality of this challenging channel model, it is not surprising that a number of open questions remain: capacity is missing in a parameter regime of the symmetric LDA which has typically been the most challenging one for the interference channel as well (the moderately weak interference regime). Constant gap to capacity results for the corresponding regime in the Gaussian channel are also missing and are an interesting topic for further investigation.

\appendix

\subsection{Proof of Theorem \ref{thm:outer bound:inj}}
\label{proof{thm:outer bound:inj}}
Given the random variables $(Q,X_1,X_2,X_0,V_1,V_2,Y_1,Y_2)$ with 
\begin{align*}
&\mathbb{P}_{Q,X_1,X_2,X_0,V_1,V_2,Y_1,Y_2}(q,x_1,x_2,x_0,v_1,v_2,y_1,y_2)
\\&=
\mathbb{P}_{Q}(q)
\mathbb{P}_{X_1|Q}(x_1|q)
\mathbb{P}_{X_2|Q}(x_2|q)
\mathbb{P}_{X_0|Q,X_1,X_2}(x_0|q,x_1,x_2)
\mathbb{P}_{V_1|X_1}(v_1|x_1)
\mathbb{P}_{V_2|X_2}(v_2|x_2)
\\& \cdot \delta\big(y_1 - f_{1}(x_1,x_0,v_2)\big)
 \delta\big(y_2 - f_{2}(x_2,x_0,v_1)\big).
 \end{align*}
let $\widetilde{V}_{1}$ and $\widetilde{V}_{2}$ be conditionally
independent copies of $V_1$ and $V_2$, distributed jointly with
$(Q,X_1,X_2,X_0)$  as
$\mathbb{P}_{\widetilde{V}_{1},\widetilde{V}_{2}|Q,X_1,X_2,X_0}(v_1,v_2|q,x_1,x_2,x_0)
=
\mathbb{P}_{V_1|X_1}(v_1|x_1)
\mathbb{P}_{V_2|X_2}(v_2|x_2).$
By Fano's inequality $H(W_i|Y_i^n)\leq n\epsilon_n, i\in[1:2],$ such that $\epsilon_n\to0$ as $n\to\infty$.
Similar arguments to those in \cite{telatar_tse} yield: 
{
\begin{align}
  &n(R_1+R_2-2\epsilon_n) 
\leq I(W_1; Y_1^n,\widetilde{V}_{1}^n)  + I(W_2; Y_2^n,\widetilde{V}_{2}^n)\notag
\\&= H(\widetilde{V}_{1}^n)
   - H(\widetilde{V}_{1}^n|W_1,X_1^n)
   + H(Y_1^n|\widetilde{V}_{1}^n) 
   - H(Y_1^n|\widetilde{V}_{1}^n,W_1,X_1^n) \notag
\\&+ H(\widetilde{V}_{2}^n)
   - H(\widetilde{V}_{2}^n|W_2,X_2^n) 
   + H(Y_2^n|\widetilde{V}_{2}^n)
   - H(Y_2^n|\widetilde{V}_{2}^n,W_2,X_2^n)\notag
\\& \stackrel{\rm (a)}{\leq}
     H(\widetilde{V}_{1}^n)
   - H(\widetilde{V}_{1}^n|X_1^n)
   + H(Y_1^n|\widetilde{V}_{1}^n)
   - H(Y_1^n|\widetilde{V}_{1}^n,W_1,X_1^n,X_0^n)\notag
\\&+ H(\widetilde{V}_{2}^n)
   - H(\widetilde{V}_{2}^n|X_2^n)
   + H(Y_2^n|\widetilde{V}_{2}^n)
   - H(Y_2^n|\widetilde{V}_{2}^n,W_2,X_2^n,X_0^n)\notag
\\
&\stackrel{\rm (b)}{=}
        H(Y_1^n|\widetilde{V}_{1}^n)
      + H(Y_2^n|\widetilde{V}_{2}^n)
      - H(\widetilde{V}_{1}^n|X_1^n)
      - H(\widetilde{V}_{2}^n|X_2^n) \notag
\\&   + H(\widetilde{V}_{1}^n)
      - H(V_1^n|W_2,\widetilde{V}_{2}^n,X_2^n,X_0^n) \notag
\\&   + H(\widetilde{V}_{2}^n)
      - H(V_2^n|W_1,\widetilde{V}_{1}^n,X_1^n,X_0^n) \notag
\\&\stackrel{\rm (c)}{=}
        H(Y_1^n|\widetilde{V}_{1}^n)
     + H(Y_2^n|\widetilde{V}_{2}^n)
      - H(\widetilde{V}_{1}^n|X_1^n)
      - H(\widetilde{V}_{2}^n|X_2^n) \notag
\\&+ H(\widetilde{V}_{1}^n) - H(V_1^n|W_2,X_0^n) 
   + H(\widetilde{V}_{2}^n) - H(V_2^n|W_1,X_0^n) \notag
\\&=    H(Y_1^n|\widetilde{V}_{1}^n) + H(Y_2^n|\widetilde{V}_{2}^n)
      - H(\widetilde{V}_{1}^n|X_1^n)
      - H(\widetilde{V}_{2}^n|X_2^n)\notag
\\&   + I(V_1^n;X_0^n|W_2)
      + I(V_2^n;X_0^n|W_1),\notag
\end{align}
}
where:
 the inequality in (a) follows from further conditioning on $X_0$ (and because given conditioning on $X_i^n$ we have that $V_i^n$ is independent of everything else, so that in particular we can drop the message $W_i$ from the conditioning, $i=1,2$),
 the equality in (b) follows from  the assumed determinism,
 the equality in (c) follows  since $V_1^n$ is independent of $(\widetilde{V}_{2}^n,X_2^n)$ so it can be dropped from the conditioning (however $X_0^n$ depends on $(W_1,W_2)$ so we must keep the messages in the conditioning) and similarly for user 2. Similarly,
%
%
{
\begin{align*}
  &n(2R_1 + R_2-3\epsilon_n)
\\& \leq
         I(W_1; Y_1^n,\widetilde{V}_{1}^n|W_2)
       + I(W_1; Y_1^n) + I(W_2; Y_2^n,\widetilde{V}_{2}^n)
\\& =    H(Y_1^n   | W_2,\widetilde{V}_{1}^n,\ X_2^n)
       - H(Y_1^n   | W_1,W_2,\widetilde{V}_{1}^n,\ X_1^n,X_2^n,X_0^n)
\\&    + H(Y_1^n)
       - H(Y_1^n   | W_1,\ X_1^n)
\\&    + H(Y_2^n   | \widetilde{V}_{2}^n)
       - H(Y_2^n   | W_2, \widetilde{V}_{2}^n,X_2^n)
\\&    + H(\widetilde{V}_{1}^n| W_2,\ X_2^n)
       - H(\widetilde{V}_{1}^n| W_1,W_2,\ X_1^n,X_2^n,X_0^n)
\\&    + H(\widetilde{V}_{2}^n  )
       - H(\widetilde{V}_{2}^n| W_2,X_2^n)
\\
&\stackrel{\rm (a)}{\leq}
     H(Y_1^n   | \widetilde{V}_{1}^n,X_2^n)
   - H(Y_1^n   | \widetilde{V}_{1}^n,X_1^n,X_2^n,X_0^n)
\\&+ H(Y_1^n)
   - H(Y_1^n   | W_1,\ X_1^n, \ X_0^n)
\\&+ H(Y_2^n   | \widetilde{V}_{2}^n)
   - H(Y_2^n   | W_2, \widetilde{V}_{2}^n,X_2^n, \ X_0^n)
\\&+ H(\widetilde{V}_{1}^n)
   - H(\widetilde{V}_{1}^n|X_1^n)
   + H(\widetilde{V}_{2}^n  )
   - H(\widetilde{V}_{2}^n| X_2^n)
\\
&\stackrel{\rm (b)}{=}
         H(Y_1^n   | \widetilde{V}_{1}^n,X_2^n)
       - H(V_2^n| \widetilde{V}_{1}^n,X_1^n,X_2^n,X_0^n)
  \\ & + H(Y_1^n)
       - H(V_2^n| W_1, X_1^n, X_0^n)
  \\ & + H(Y_2^n| \widetilde{V}_{2}^n)
       - H(V_1^n| W_2, \widetilde{V}_{2}^n,X_2^n, X_0^n)
  \\ & + H(\widetilde{V}_{1}^n)
       - H(\widetilde{V}_{1}^n|X_1^n)
       + H(\widetilde{V}_{2}^n  )
       - H(\widetilde{V}_{2}^n| X_2^n)\\
&\stackrel{\rm (c)}{=}
         H(Y_1^n   | \widetilde{V}_{1}^n,X_2^n)
       - H(V_2^n| X_2^n)
\\ &+ H(Y_1^n)
    - H(V_2^n| W_1, X_0^n)
\\ &+ H(Y_2^n| \widetilde{V}_{2}^n)
    - H(V_1^n| W_2, X_0^n)
\\ & + H(\widetilde{V}_{1}^n)
    - H(\widetilde{V}_{1}^n|X_1^n)
    + H(\widetilde{V}_{2}^n  )
    - H(\widetilde{V}_{2}^n| X_2^n)\\
 &\leq H(Y_1^n)
       + H(Y_1^n   | \widetilde{V}_{1}^n,X_2^n)
       + H(Y_2^n   | \widetilde{V}_{2}^n)\\ & - H(\widetilde{V}_{1}^n|X_1^n)
       - 2H(V_2^n| X_2^n)
       + I(V_2^n; X_0^n|W_1)
       + I(V_1^n; X_0^n|W_2),
\end{align*}
}
where the inequalities labeled (a), (b) and (c) follow from the same reasoning
used in the  in the derivation of the sum-rate bound.
The remaining bound  is obtained by swapping the users.

\subsection{Proof of Theorem \ref{thm:outer bound:LDA}}
\label{proof{thm:outer bound:LDA}}
For the channels in~\eqref{eq:further restriction of injective semidet to fit LDA}, instead of conditioning on $X_0$ in the step marked by (a) in Appendix~\ref{proof{thm:outer bound:inj}},  we condition on the $q_i(X_0), \ i\in[1,2],$
to obtain the tighter bound
{
\begin{align*}
  & H(\widetilde{V}_{2}^n)-H(Y_1^n|\widetilde{V}_{1}^n,W_1,X_1^n)\notag
\\&\stackrel{\rm (a')}{\leq} H(\widetilde{V}_{2}^n)-H(Y_1^n|\widetilde{V}_{1}^n,W_1,X_1^n, \ q_1(X_0^n)) \notag
\\&= H(V_2^n)-H(V_2^n|\widetilde{V}_{1}^n,W_1,X_1^n,q_1(X_0^n)) \notag
 = H(V_2^n)-H(V_2^n|W_1,q_1(X_0^n)) \notag
\\&= I(V_2^n;W_1,q_1(X_0^n)) 
   = I(V_2^n;q_1(X_0^n)|W_1) \notag
\\&\leq  \min\{ H(V_2^n),H(q_1(X_0^n))\}
\leq  n\min\{ H(V_2|Q),H(q_1(X_0)|Q)\} ,\notag
\end{align*}
}
and similarly for the other users.
The fact that the resulting region is exhausted by  i.i.d. Bernoulli($1/2$) bits for the input vectors follows by arguments similar to~\cite{avestimehr2011wireless}.

{

\subsection{Proof of Theorem \ref{thm:outer bound:Gaussian}}
\label{proof{thm:outer bound:Gaussian}}

Inspired by the proof of Theorem~\ref{thm:outer bound:LDA} --- where the term $H(Y_i^n|\widetilde{V}_{i}^n,W_i,X_i^n), \ i\in[1:2],$ was further conditioned on $q_i(X_0^n)$ rather than on $X_0^n$ (i.e., compare step marked by (a) in Appendix~\ref{proof{thm:outer bound:inj}} with step marked by (a') in Appendix~\ref{proof{thm:outer bound:LDA}}) --- we mimic here the function $q_i(X_0^n)$ for the LDA with $|h_{i0}|X_0^n-Z_0^n$ for the GICCR, where
$Z_0 \ \text{i.i.d.} \ \mathcal{N}(0,1)$
independent of
$(Z_1,Z_2,\widetilde{Z}_1,\widetilde{Z}_2,W_1,W_2).$
Recall that 
\begin{align*}
  &V_2 = h_{12} X_2+Z_1 \sim \widetilde{V}_{2} = h_{12} X_2+\widetilde{Z}_1 :
Z_2 \ \text{independent of} \  \widetilde{Z}_1 \sim Z_1,
\\&V_1 = h_{21} X_1+Z_2 \sim \widetilde{V}_{1} = h_{21} X_1+\widetilde{Z}_2 :
Z_1 \ \text{independent of} \  \widetilde{Z}_2 \sim Z_2.
\end{align*}
Then, we replace the step marked with (a) in Appendix~\ref{proof{thm:outer bound:inj}} with
\begin{align*}
  &      h(\widetilde{V}_{2}^n) - h(Y_1^n|\widetilde{V}_{1}^n,W_1,X_1^n)
\\&\stackrel{\rm (a')}{\leq}  h(h_{12} X_2^n+\widetilde{Z}_1^n) - h(|h_{11}|X_1^n+|h_{10}|X_0^n+ h_{12} X_2^n+Z_1^n|h_{21} X_1^n+\widetilde{Z}_{2}^n,W_1,X_1^n, \ |h_{10}|X_0^n-Z_0^n)
\\&=  h(h_{12} X_2^n+Z_1^n) - h(h_{12} X_2^n+Z_1^n+Z_0^n|W_1, |h_{10}|X_0^n-Z_0^n)
\\&=   - I(h_{12} X_2^n+Z_1^n+Z_0^n; Z_0^n)
       + I(h_{12} X_2^n+Z_1^n+Z_0^n; |h_{10}|X_0^n-Z_0^n | W_1)
\\&\leq  -0 + h(|h_{10}|X_0^n-Z_0^n) - h(|h_{10}|X_0^n-Z_0^n| W_1, h_{12} X_2^n+Z_1^n+Z_0^n, \ W_2)
\\&=  I(|h_{10}|X_0^n-Z_0^n;X_0^n) +I(Z_0^n; Z_1^n+Z_0^n)
 \leq  n\log(1+|h_{10}|^2) + n\log(2).
\end{align*}
We can also trivially upper bound $\mathsf{MLP}_1$ in~\eqref{eq:outer bound:inj:mlp1} as
\begin{align*}
  &      h(\widetilde{V}_{2}^n) - h(Y_1^n|\widetilde{V}_{1}^n,W_1,X_1^n)
   \leq I(V_2^n;X_0^n | W_1) \quad \text{as per Theorem~\ref{thm:outer bound:inj}}
\\&\leq h(h_{12} X_2^n+Z_1^n)- h(h_{12} X_2^n+Z_1^n|X_0^n,X_1^n,W_1, \ W_2)
\\&= I(h_{12}X_2^n+Z_1^n;X_2^n)
   \leq n\log(1+|h_{12}|^2).
\end{align*}
Therefore, we conclude that 
\begin{align*}
\frac{h(\widetilde{V}_{2}^n) - h(Y_1^n|\widetilde{V}_{1}^n,W_1,X_1^n)}{n} \leq
\log(1+\min\{|h_{12}|^2,|h_{10}|^2\})+\log(2).
\end{align*}
By repeating the same reasoning for the other receiver,  we conclude that for the GICCR Theorem~\ref{thm:outer bound:inj} holds with $\mathsf{MLP}_1$ in~\eqref{eq:outer bound:inj:mlp1} replaced by
\begin{align*}
\mathsf{MLP}_2
:=
  \log(1+\min\{|h_{12}|^2,|h_{10}|^2\})
 +\log(1+\min\{|h_{21}|^2,|h_{20}|^2\})
 +2\log(2).
\end{align*}
The resulting region is exhausted by jointly Gaussian inputs by arguments similar to~\cite{riniIFCCR}.

\subsection{Evaluation of the sum-rate upper bound in~\eqref{eq:outer bound:inj sumrate} for the GICCR}
\label{elavulationofETWsumrate}
By Theorem~\ref{thm:outer bound:Gaussian} we can restrict attention to jointly Gaussian inputs.
Let parameterize the possible jointly Gaussian inputs as
\begin{align*}
  &\begin{bmatrix} X_1 \\ X_2 \\ X_0 \\ \end{bmatrix}
  \sim \mathcal{N}\left( 0,  \begin{bmatrix} 1 & 0 & r_1^* \\ 0 & 1 & r_2^* \\ r_1 & r_2 & 1 \\ \end{bmatrix} \right) : |r_1|^2 + |r_2|^2 \leq 1,
\end{align*}
that is, $X_0 = r_1 X_1 + r_2 X_2 + X_0'$ with $X_0' \sim \mathcal{N}(0, 1- |r_1|^2 - |r_2|^2)$ and independent of everything else.
In~\eqref{eq:outer bound:inj sumrate}, consider the term
\begin{align*}
  &h(Y_1|\widetilde{V}_1,Q) - h(\widetilde{V}_2|X_2)
   \leq h(|h_{11}|X_1+|h_{10}|X_0+ h_{12} X_2+Z_1|h_{21} X_1 + \tilde{Z}_2) - h(\tilde{Z}_1)
\\&= h(a_1 X_1+|h_{10}|X_0'+ a_2 X_2+Z_1|h_{21} X_1 + \tilde{Z}_2)|_{a_1 := |h_{11}| + r_1|h_{10}|, \ a_2 :=  h_{12}  + r_2|h_{10}|} - h(Z_1)
\\&= \log\left(\frac{|a_1|^2}{1+|h_{21}|^2} +|h_{10}|^2(1- |r_1|^2 - |r_2|^2)+ |a_2|^2+1\right)|_{a_1 := |h_{11}| + r_1|h_{10}|, \ a_2 :=  h_{12}  + r_2|h_{10}|}
\\&\leq \log\left(\frac{(|h_{11}| + |r_1||h_{10}|)^2}{1+|h_{21}|^2} +|h_{10}|^2(1- |r_1|^2 - |r_2|^2)+ (|h_{12}|  + |r_2||h_{10}|)^2+1\right)
\\&= \log\left(\frac{(|h_{11}| + |r_1||h_{10}|)^2}{1+|h_{21}|^2} +|h_{10}|^2(1- |r_1|^2)  + 2|r_2||h_{10}||h_{12}|+ |h_{12}|^2+1\right)
\end{align*}
where clearly the last expression, for any $r_1$ such that $1- |r_1|^2\geq 0$, is maximized by $|r_2|=\sqrt{ 1-|r_1|^2}$  (recall that the bound must be optimized over $|r_1|^2 + |r_2|^2 \leq 1$); this implies that for some $|r_1|^2 + |r_2|^2 = 1$
\begin{align}
  &h(Y_1|\widetilde{V}_1,Q) - h(\widetilde{V}_2|X_2)
\leq \log\left(\frac{(|h_{11}| + |r_1||h_{10}|)^2}{1+|h_{21}|^2} + (|h_{12}|  + |r_2||h_{10}|)^2+1\right).
\label{eq:app:1}
\end{align}
By a similar reasoning for the other receiver, we have that for some $|r_1|^2 + |r_2|^2 = 1$
\begin{align}
h(Y_2|\widetilde{V}_2,Q)-h(\widetilde{V}_1|X_1)
\leq \log\left((|h_{21}| + |r_1||h_{20}|)^2  + \frac{(|h_{22}| + |r_2||h_{20}|)^2}{1+|h_{12}|^2}+1\right).
\label{eq:app:2}
\end{align}
Finally, by summing~\eqref{eq:app:1} and~\eqref{eq:app:2}, the sum-rate upper bound from Theorem~\ref{thm:outer bound:inj} with the MLP from Theorem~\ref{thm:outer bound:Gaussian} reads
\begin{align}
  R_1+R_2
  &\leq \max_{|r_1|^2 + |r_2|^2 = 1}
    \log\left(\frac{(|h_{11}| + |r_1||h_{10}|)^2}{1+|h_{21}|^2} + (|h_{12}|  + |r_2||h_{10}|)^2+1\right)
\notag\\&+\log\left((|h_{21}| + |r_1||h_{20}|)^2  + \frac{(|h_{22}| + |r_2||h_{20}|)^2}{1+|h_{12}|^2}+1\right)
\notag\\&+\log(1+\min\{|h_{12}|^2,|h_{10}|^2\})
+\log(1+\min\{|h_{21}|^2,|h_{20}|^2\})
+2\log(2).
\label{eq:app:3}
\end{align}

In the symmetric case in~\eqref{eq:GICCR snr exponents interf}, by the symmetry of the problem, it is easy to see that the maximizing $(r_1,r_2)$ is such that $|r_1|^2 = |r_2|^2 = 1/2$; hence the sum-rate upper bound in~\eqref{eq:app:3} reads
\begin{align}
  R_1+R_2
&\leq
    2\log\left(\frac{ \max\{|h_{\rm{S}}|^2,|h_{\rm{C}}|^2\}}{1+|h_{\rm{I}}|^2} + \max\{|h_{\rm{I}}|^2,|h_{\rm{C}}|^2\}+1\right)
\notag\\&+2\log(1+\min\{|h_{\rm{I}}|^2,|h_{\rm{C}}|^2\})
+2\log(2(1 + 1/\sqrt{2})^2)
\label{eq:etw sym summate for sic regime}
\end{align}
where
$2\log(2(1 + 1/\sqrt{2})^2) \leq 2\log(6) < 5.17$~bits.

\subsection{Lower bounds on the Achievable Rates for the Scheme in Section~\ref{sec:VI.1gap}}
\label{app:VI.1gap}
The achievable scheme in Section~\ref{sec:VI.1gap} attains the following rates (where the further lower bonds follow from straightforward but tedious algebraic manipulations by using the conditions (c1)-(c5) of Theorem~\ref{thm:an achievable region in Regime VI.1})
\begin{subequations}
\begin{align}
R_{A_1}&\leq \log \left(1+\frac{|h_{\rm{S}}|^2 \left(\frac{|h_{\rm{I}}|^2}{1+|h_{\rm{I}}|^2}-\frac{|h_{\rm{C}}|^2}{2|h_{\rm{I}}|^2}\right)}{1+\frac{|h_{\rm{S}}|^2}{1+|h_{\rm{I}}|^2}+\frac{|h_{\rm{I}}|^2}{1+|h_{\rm{I}}|^2} +\frac{|h_{\rm{C}}|^2}{2}\  \left|\frac{|h_{\rm{S}}|}{|h_{\rm{I}}|}{\rm e}^{-{\rm j}\angle{h_{21}}}-1\right|^2 } \right)
\label{eq:achievable:rateA1 at Y1}
\\& :
\frac{|h_{\rm{S}}|^2 \left(\frac{|h_{\rm{I}}|^2}{1+|h_{\rm{I}}|^2}-\frac{|h_{\rm{C}}|^2}{2|h_{\rm{I}}|^2}\right)}{1+\frac{|h_{\rm{S}}|^2}{1+|h_{\rm{I}}|^2}+\frac{|h_{\rm{I}}|^2}{1+|h_{\rm{I}}|^2} +\frac{|h_{\rm{C}}|^2}{2}\  \left|\frac{|h_{\rm{S}}|}{|h_{\rm{I}}|}{\rm e}^{-{\rm j}\angle{h_{21}}}-1\right|^2 } \geq \frac{|h_{\rm{I}}|^2}{4} \frac{1}{3+|h_{\rm{C}}|^2},
\notag
\\
R_{A_2}&= \log \left(1+\frac{|h_{\rm{C}}|^2}{2}\ \frac{\left|\frac{|h_{\rm{S}}|}{|h_{\rm{I}}|}{\rm e}^{-{\rm j}\angle{h_{21}}}-1\right|^2}{1+\frac{|h_{\rm{S}}|^2}{1+|h_{\rm{I}}|^2}+\frac{|h_{\rm{I}}|^2}{1+|h_{\rm{I}}|^2}}\right)
\geq \left. \log\left(1+|h_{\rm{C}}|^2\frac{(\sqrt{t}-1)^2}{2(2+t)} \right)\right|_{t:=\frac{|h_{\rm{S}}|^2}{1+|h_{\rm{I}}|^2}}
\notag\\&\geq \log \left(1+\frac{|h_{\rm{C}}|^2}{n_{A_2}} \right) \ \text{for $\sqrt{t}\geq \frac{1+\sqrt{4/n_{A_2}(3/2-2/n_{A_2})}}{1-2/n_{A_2}}$},
\label{eq:achievable:rateA2}
\\
R_{A_3}&= \log\left(1+\frac{\frac{|h_{\rm{S}}|^2}{1+|h_{\rm{I}}|^2}}{1+\frac{|h_{\rm{I}}|^2}{1+|h_{\rm{I}}|^2}}\right)
        = \log\left(1+\frac{|h_{\rm{S}}|^2}{1+2|h_{\rm{I}}|^2}\right),
\label{eq:achievable:rateA3}
\\
R_{B_1}&= \log\left(1+\frac{|h_{\rm{C}}|^2}{2} \ \frac{\left|\frac{|h_{\rm{S}}|}{|h_{\rm{I}}|}{\rm e}^{-{\rm j}\angle{h_{12}}}-1 \right|^2}{1+|h_{\rm{I}}|^2\left(\frac{|h_{\rm{I}}|^2}{1+|h_{\rm{I}}|^2}-\frac{|h_{\rm{C}}|^2}{2|h_{\rm{I}}|^2}\right)+\frac{|h_{\rm{I}}|^2}{1+|h_{\rm{I}}|^2}+\frac{|h_{\rm{S}}|^2}{1+|h_{\rm{I}}|^2}}\right)
\geq \left. \log\left(1+|h_{\rm{C}}|^2\frac{(\sqrt{t}-1)^2}{4(1+t)} \right) \right.
\notag\\&\geq \log \left(1+\frac{|h_{\rm{C}}|^2}{n_{B_1}} \right) \ \text{for $\sqrt{t}\geq \frac{1+\sqrt{8/n_{B_1}(1-2/n_{B_1})}}{1-4/n_{B_1}}$},
\label{eq:achievable:rateB1}
\\
R_{B_2}&= \log\left(1+\frac{|h_{\rm{S}}|^2\left(\frac{1}{1+|h_{\rm{I}}|^2}-\frac{|h_{\rm{C}}|^2}{1+|h_{\rm{S}}|^2}\right)}{1+|h_{\rm{I}}|^2\left(\frac{|h_{\rm{I}}|^2}{1+|h_{\rm{I}}|^2}-\frac{|h_{\rm{C}}|^2}{2|h_{\rm{I}}|^2}\right)+\frac{|h_{\rm{I}}|^2}{1+|h_{\rm{I}}|^2}+\frac{|h_{\rm{S}}|^2|h_{\rm{C}}|^2}{1+|h_{\rm{S}}|^2}}\right)
\geq \log \left(1+\frac{|h_{\rm{S}}|^2}{4(1+|h_{\rm{I}}|^2)^2} \right),
\label{eq:achievable:rateB2}
\\
R_{A_1}&\leq \log\left(1+\frac{|h_{\rm{I}}|^2\left(\frac{|h_{\rm{I}}|^2}{1+|h_{\rm{I}}|^2}-\frac{|h_{\rm{C}}|^2}{2|h_{\rm{I}}|^2} \right)}{1+\frac{|h_{\rm{I}}|^2}{1+|h_{\rm{I}}|^2}+\frac{|h_{\rm{S}}|^2|h_{\rm{C}}|^2}{1+|h_{\rm{S}}|^2}} \right)
\label{eq:achievable:rateA1 at Y2}
\quad
: \frac{|h_{\rm{I}}|^2\left(\frac{|h_{\rm{I}}|^2}{1+|h_{\rm{I}}|^2}-\frac{|h_{\rm{C}}|^2}{2|h_{\rm{I}}|^2} \right)}{1+\frac{|h_{\rm{I}}|^2}{1+|h_{\rm{I}}|^2}+\frac{|h_{\rm{S}}|^2|h_{\rm{C}}|^2}{1+|h_{\rm{S}}|^2}} \geq \frac{|h_{\rm{I}}|^2}{4(2+|h_{\rm{C}}|^2)},
\\
R_{B_3} &= \log\left(1+\frac{|h_{\rm{C}}|^2 \ \frac{|h_{\rm{S}}|^2}{1+|h_{\rm{S}}|^2}}{1+\frac{|h_{\rm{I}}|^2}{1+|h_{\rm{I}}|^2}}\right) \geq \log\left(1+\frac{|h_{\rm{C}}|^2}{4}\right). 
\label{eq:achievable:rateB3}
\end{align}
and, because $X_{A_1}$ is a ``common message'' decoded at both destinations, we finally choose
\begin{align}
R_{A_1} = \min\{\text{eq.\eqref{eq:achievable:rateA1 at Y1}, eq.\eqref{eq:achievable:rateA1 at Y2}}\}
\geq \log\left(1+\frac{|h_{\rm{I}}|^2}{4(3+|h_{\rm{C}}|^2)}\right).
\label{eq:achievable:rateA1}
\end{align}
\label{eq:achievable:rateABlower}
\end{subequations}

As outer bound consider the corner point obtained by intersecting the sum-rate upper
bound in~\eqref{eq:etw sym summate for sic regime}
with the single-rate upper
bound in~\eqref{eq:out:DM1} (see eq.\eqref{eq:piecewise:outerbound})
whose coordinates are given in~\eqref{eq:ourterboud:CornerPont2mainall} (note that in this regime the channel gains satisfy $|h_{\rm{C}}|^2 \leq |h_{\rm{I}}|^2 \leq |h_{\rm{S}}|^2$).
We next compare the lower bound in~\eqref{eq:achievable:rateABlower} with the corner point outer bound in~\eqref{eq:ourterboud:CornerPont2mainall}.
It can be easily seen that the gap for $R_1$ is, for $n_{A_2}\geq 3, n_{B_1}\geq 4$,
\begin{align}
\text{gap}_{R_1}
&\leq \text{eq.\eqref{eq:ourterboud:CornerPont1} - eq.\eqref{eq:achievable:rateA1} - eq.\eqref{eq:achievable:rateA2} - eq.\eqref{eq:achievable:rateA3}}
\notag\\&= \log \frac{(1+4 |h_{\rm{S}}|^2)(3+|h_{\rm{C}}|^2)(1+2 |h_{\rm{I}}|^2)}
{(1+ |h_{\rm{S}}|^2+2 |h_{\rm{I}}|^2)(1+|h_{\rm{C}}|^2/n_{A_2})(3+|h_{\rm{C}}|^2+|h_{\rm{I}}|^2/4)}
\leq \log(4\cdot n_{A_2} \cdot8), 
\label{eq:gapVI.1.R1}
\\
\text{gap}_{R_2}
  &\leq \text{eq.\eqref{eq:ourterboud:CornerPont2} - eq.\eqref{eq:achievable:rateB1} - eq.\eqref{eq:achievable:rateB2} - eq.\eqref{eq:achievable:rateB3}}
\notag\\&= \log \frac{4(1 + 1/\sqrt{2})^4\left( 1+|h_{\rm{I}}|^2+\frac{|h_{\rm{S}}|^2}{1+|h_{\rm{I}}|^2} \right)^2(1+|h_{\rm{C}}|^2)^2}{(1+4|h_{\rm{S}}|^2)\left(1+\frac{|h_{\rm{S}}|^2}{4(1+|h_{\rm{I}}|^2)^2} \right)(1+|h_{\rm{C}}|^2/n_{B_1})(1+|h_{\rm{C}}|^2/4)}
\notag\\&\leq 
\left. \log \frac{4(1 + 1/\sqrt{2})^4\cdot(1+2 t)^2\cdot n_{B_1}\cdot4}{1+t^2}\right|_{t:=\frac{|h_{\rm{S}}|^2}{1+|h_{\rm{I}}|^2} \geq t_0, \sqrt{t_0} := \max\left\{\frac{1+\sqrt{4/n_{A_2}(3/2-2/n_{A_2})}}{1-2/n_{A_2}},\frac{1+\sqrt{8/n_{B_1}(1-2/n_{B_1})}}{1-4/n_{B_1}}\right\}}
\notag\\&\leq \log\left(4(1 + 1/\sqrt{2})^4\cdot 5 \cdot n_{B_1}\cdot4\right).
\label{eq:gapVI.1.R2}
\end{align}
Thus the proposed scheme achieves a corner point of the capacity region outer bound to within at most
$
\text{gap}
=
\max\{\text{ eq.\eqref{eq:gapVI.1.R1}, eq.\eqref{eq:gapVI.1.R2} }\}.
$
For example, for $n_{A_2}=n_{B_1}=10$, we have $t_0=9$ and $\text{gap}=\gapVIone~\text{bits per user}.$
The gap can be reduced by increasing the value of $t_0$.

}

\bibliography{refs}
\bibliographystyle{IEEEtran}

\end{document}